\documentclass{ab}

\usepackage{graphicx}
\usepackage[numbers]{natbib}
\usepackage{multirow}
\usepackage{mathtools}
\usepackage{hyperref}
\usepackage{xcolor}
\usepackage{tabularx}

\begin{document}

\title{Large Scale Distribution of Galaxies in The Field HS\,47.5-22. II. Observational Data Analysis}

\author{A.~A. Grokhovskaya and S. ~N. Dodonov and T. ~A. ~Movsesyan}

\institute{Special Astrophysical Observatory, Russian Academy of Sciences, Nizhnij Arkhyz, 369167, Russia and Byurakan Astrophysical Observatory, Aragatsotn reg., Armenia}

 \titlerunning{Large Scale Distribution of Galaxies in The Field HS\,47.5-22. II. Observational Data Analysis}

\authorrunning{Grokhovskaya}

\date{May 19,  2020/Revised: ~}
\offprints{A. Grokhovskaya  \email{grohovskaya.a@gmail.com} }

\abstract{
The results of the study of the large-scale distribution of galaxies up to $ z \sim 0.8 $ in the field HS 47.5-22 on the basis of photometric data of the 1-meter Schmidt telescope of the BAO NAS Armenia are presented. The full sample contains 28 398 galaxies. Candidates for large-scale structures were determined using two independent methods for reconstructing density contrast maps in 57 narrow slices of the three-dimensional distribution of galaxies: adaptive aperture algorithm with smoothing and 2D Voronoi tessellation. We have identified more than 250 statistically significant overdense structures. The obtained results demonstrate a wide range of overdense structures over the full range of redshifts up to $ z \sim 0.8 $.
}

\maketitle

\section{Introduction}
\label{intro}

The evolution and physical properties of galaxies are closely related to the properties of their environment. The dependence of the morphology of galaxies on the density of the environment was first discovered in \citep{Oemler1974, Dressler1980}. Later it turned out that not only morphology but also other physical properties of galaxies correlate with the density of the environment. Thus, local density affects color excesses, the equivalent line width $ H_ \alpha $ and the magnitude of the jump D4000 ~\AA \citep{Kauffmann2004}, the star formation rate for galaxies at close redshifts \citep{Peng2010}. Studies based on 2dFGRS \citep[2dF Galaxy Redshift Survey,][]{Madgwick2003} and SDSS \citep[Sloan Digital Sky Survey,][]{Guo2013, Guo2014} surveys showed that the relationship between the local environment and morphology. It is presented not only in clusters of galaxies but exists for the entire range of local densities up to field galaxies.

The use of spectroscopic redshifts is most preferable for analyzing the large-scale distribution of galaxies and the dependence of their physical properties on the density of the environment. Spectroscopic redshifts were widely used in research on small z with relatively bright galaxies (e.g. in \citep{Peng2010}). However, for samples of tens and hundreds of thousands of galaxies with high redshifts, fainter than $ I_{AB} = 22^m $ and without strong emission lines, this is practically impossible. Spectroscopy of such faint galaxies requires the largest telescopes and exposure times of several hours \citep {LeFevre2005, Gerke2005, Meneux2006, Cooper2006, Coil2007, Lilly2007}. That is why photometric surveys using mid-band filters are becoming increasingly relevant.

There are only a few surveys of sufficient depth more or less suitable for solving the problem of analyzing the large-scale distribution of galaxies: COMBO-17 \citep [Classifying Objects by Medium-Band Observations, a spectrophotometric 17-filter survey,] [] {Wolf2004}, ALHAMBRA \cite [Advanced Large, Homogeneous Area Medium Band Redshift Astronomical Survey,] [] {Moles2008}, COSMOS \citep [Cosmic Evolution Survey,] [] {Murayama2007}. Some of these surveys, despite the significant total area (for COMBO-17 - $0.78  \; \mathrm {deg ^ 2}$ , for ALHAMBRA - $2.79  \; \mathrm {deg ^ 2} $), were performed on small-sized areas significantly spaced in space, which does not allow fully recovering the large-scale distribution of galaxies. Broadband surveys, due to the low accuracy of determining photometric redshifts and spectral type classification of galaxies, are excluded from consideration. Spectral surveys of sufficient area (\citep [Sloan Digital Sky Survey,] [] {Peng2010}, 2dFGRS) are limited in-depth, while deep spectral surveys are insufficient in the area and sample samples are not complete due to the need for preliminary selection of objects.

The most successful survey - COSMOS  has a total area of $ 1.95 \; \mathrm {deg ^ 2} $ \cite {Murayama2007} and represents a single field (coordinates of the center $ 10 ^ h00 ^ m28.60 ^ s \; + 02 ^ d12 ^ m21.0 ^ s $), uniformly covered by observations at 30 filters. In addition, spectral information was obtained for nearly 20,000 galaxies in this field. Based on these data, \cite {Scoville2013} investigated the large-scale distribution of galaxies in this field up to the redshift $ z \sim 3.0 $ using Voronoi tessellation diagrams and adaptive smoothing. The authors of this work confirm the dependence of the physical parameters of galaxies (stellar mass, spectral energy distribution, star formation rate) on the density of the environment, and also show a strong dependence of the position for early-type galaxies on the dense regions of the large-scale distribution of galaxies.

This work presents the results of a study of the large-scale distribution of the galaxy density of field HS47.5 - 22 using high-precision photometric redshifts of up to $ z \sim 0.8 $ and qualitative classification of galaxies by types of spectral energy distribution. Section 2 describes the observations. Section 3 discusses the selection of galaxies, the determination of photometric redshifts, and the classification of galaxies. Two independent methods were used to determine large-scale structures - an algorithm with an adaptive aperture and smoothing and two-dimensional Voronoi diagrams (Section 4). Maps of the large-scale distribution of galaxies, as well as an analysis of the results, are presented in Section 5. The article uses the cosmological $ \Lambda CDM $ model with parameters $ \Omega_M = 0.3 $, $ \Omega_ \Lambda = 0.7 $ and $ \mathrm {H_0 = 70 \; km \; s ^ {- 1}} $.

\section{Observations}
The observations were carried out at the 1-meter Schmidt telescope \cite{Dodonov2017} of the Byurakan Astrophysical Observatory for several sets in February, March, April and November 2018 and in February and November 2019. The field HS 47.5-22 (with center coordinates $ 09 ^ h50 ^ m00 ^ s \; + 47 ^ d35 ^ m00 ^ s $) is one of the medium deep fields of the X-ray satellite ROSAT \citep {Molthagen1997}. Total exposure time $> 5000$ sec. for $ 73 \% $ area and more than 20,000 sec. for the central region of size $ 2.3 \; \mathrm {deg ^ 2} $. Limiting flux $ {3.4 * 10 ^ {- 14} \; \mathrm {ergs \; cm ^ {- 2} \; s ^ {- 1}}} $ was obtained for objects in the energy range 0.1–2.4 keV. The choice of the field for observations is due to its location in the region with a very low density of neutral hydrogen on the line of sight $ {\mathrm {<N_H>} = 10 ^ {20} \; \mathrm {cm ^ {- 2}}} $, which is not much higher than the absorption value in the region of the "Lockman hole" \cite {Lockman1986}, where the lowest absorption on the line of sight for the northern sky is observed $ {\mathrm {<N_H>} = 4.5 * 10 ^ {19} \; \mathrm {cm ^ {- 2}}} $.

For the field HS47.5-22, observations were made in 4 broadband filters (u, g, r, i SDSS) and 16 mid-band filters (FWHM = 250  ~\AA), with a uniform coverage of the spectral range 4000 - 8000 $ \mathrm { \AA \AA} $) to $ R_\mathrm {AB} = 23 ^ m $ (see Tab. \ref {tabl1}, Fig. 1). The central part of the field, the size of $ 2.386 \; \mathrm {deg ^ 2} $, was covered by four sets of exposures in broad-band and medium-band filters. The overlap of adjacent sets was about 10 arcmin. The total exposure time was selected so as to reach a depth of $ m_{AB} \approx 25 ^ m $ with a signal-to-noise ratio of $ \sim 5 $ in broadband (about 2 hours) and $ m_{AB} \approx 23 ^ m $ with a signal-to-noise ratio of $ \sim 5 $ in mid-range filters (about 60 minutes at the peak of the detector sensitivity curve and about 2 hours at the edges of the range).

\begin{figure}[h]
\hspace*{10pt}
\includegraphics[width=1.0\linewidth]{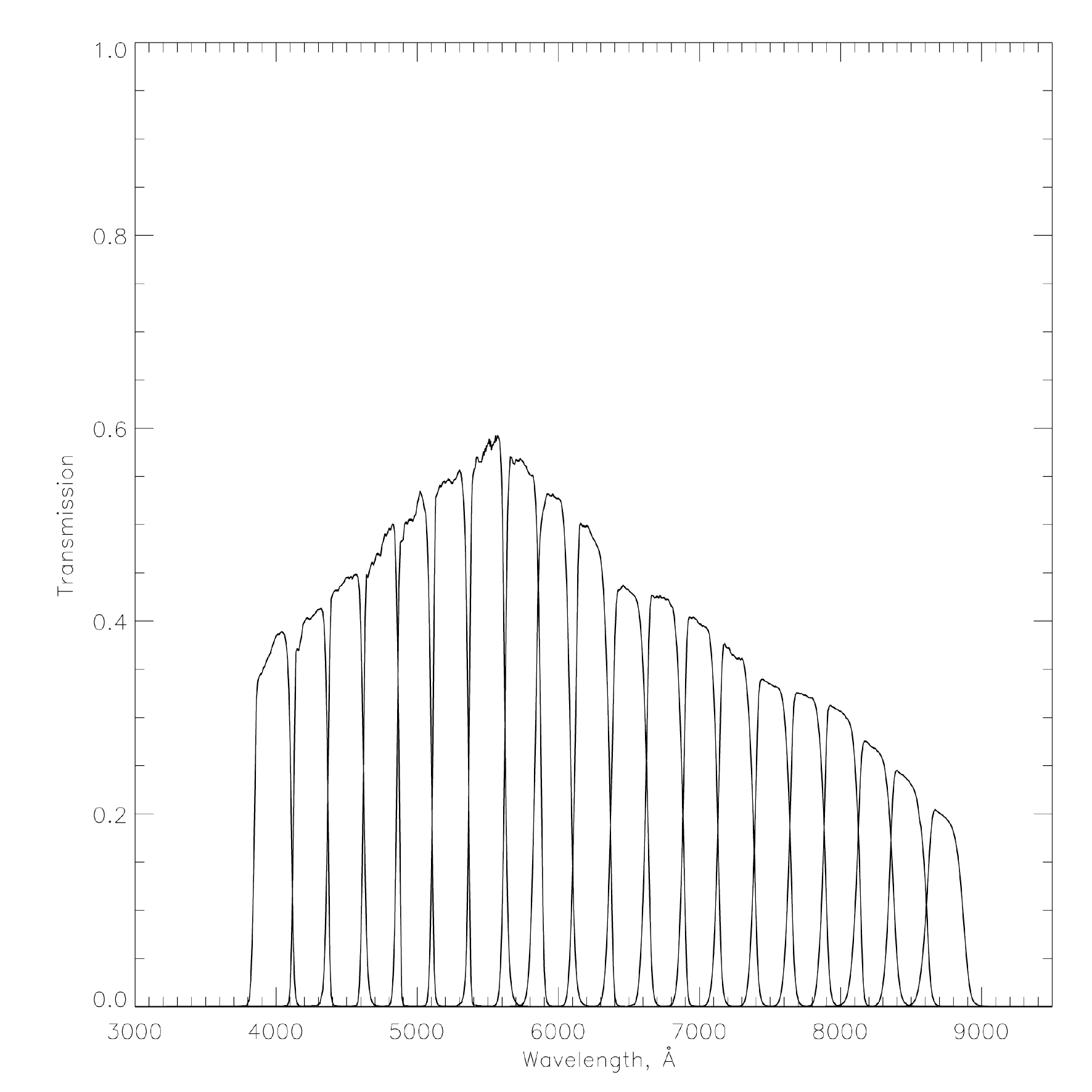}
\caption{A set of filters of the 1-meter Schmidt telescope BAO NAS. Filter transmission was measured in F / 2 the spectral sensitivity of the CCD detector were included.}
\label{fig_1}
\end{figure}

\begin{table}[ht]
\caption{A set of filters of the 1-meter  Schmidt telescope BAO NAS. Effective wavelength, FWHM, Limiting magnitude measured at signal-to-noise level $ \sim 5$.} 
\centering 
\begin{tabular}{c c c c} 
\hline\hline 
Filter & $\lambda_{\mathrm{cen}}, \AA$ & FWHM, $\AA$  & $m_{\mathrm{lim},5\sigma}$\\ [0.5ex] 
\hline 
u\_SDSS & 3578 & 338 & 24.23 \\ 
g\_SDSS & 4797 & 860 & 25.22 \\
r\_SDSS & 6227 & 770 & 24.97 \\
i\_SDSS & 7624 & 857 & 24.15 \\
MB\_400 & 3978 & 250 & 24.37 \\
MB\_425 & 4246 & 250 & 24.31 \\
MB\_450 & 4492 & 250 & 24.20 \\
MB\_475 & 4745 & 250 & 24.31 \\
MB\_500 & 4978 & 250 & 24.30 \\
MB\_525 & 5234 & 250 & 24.37 \\
MB\_550 & 5496 & 250 & 23.86 \\
MB\_575 & 5746 & 250 & 24.29 \\
MB\_600 & 5959 & 250 & 23.89 \\
MB\_625 & 6234 & 250 & 23.51 \\
MB\_650 & 6499 & 250 & 23.41 \\
MB\_675 & 6745 & 250 & 23.78 \\
MB\_700 & 7002 & 250 & 23.47 \\
MB\_725 & 7253 & 250 & 23.20 \\
MB\_750 & 7519 & 250 & 23.07 \\
MB\_775 & 7758 & 250 & 22.97 \\[1ex] 
\hline 
\end{tabular}
\label{tabl1} 
\end{table}

\section {Photometric redshifts and galaxy selection}
The galaxy sample of the HS47.5-22 field is based on deep ($ m_ {AB} \approx 25 ^ m $) images in the broad-band filters of the SDSS system ($ g $, $ r $ and $ i $) and is limited by the threshold magnitude $ m_ {AB} \approx 23 ^ m $, up to which the images in mid-band filters with signal-to-noise ratio $ \sim 5 $ were received. The total number of field galaxies is about 100,000; the sample of galaxies intended for the study included 28,398 galaxies applicable the selection criteria (see Section 3.1).

The photometric properties of a galaxy sample  of the HS 47.5-22 field were investigated in the range of 4000 ~\AA up to 8000 ~\AA, which allows you to confidently determine the redshifts of galaxies from $ z = 0 $ to $ z = 0.8 $. Photometry of galaxies was performed using the SExtractor \cite {Bertin1996} software package in dual image mode, where a composite image of the field was used as a reference image, created on the basis of images in SDSS broadband g, r and i filters (the photometry procedure is detailed and data calibration is described in \cite {Dodonov2020}).

\begin{figure}[h]
\hspace*{10pt}
\includegraphics[width=0.9\linewidth]{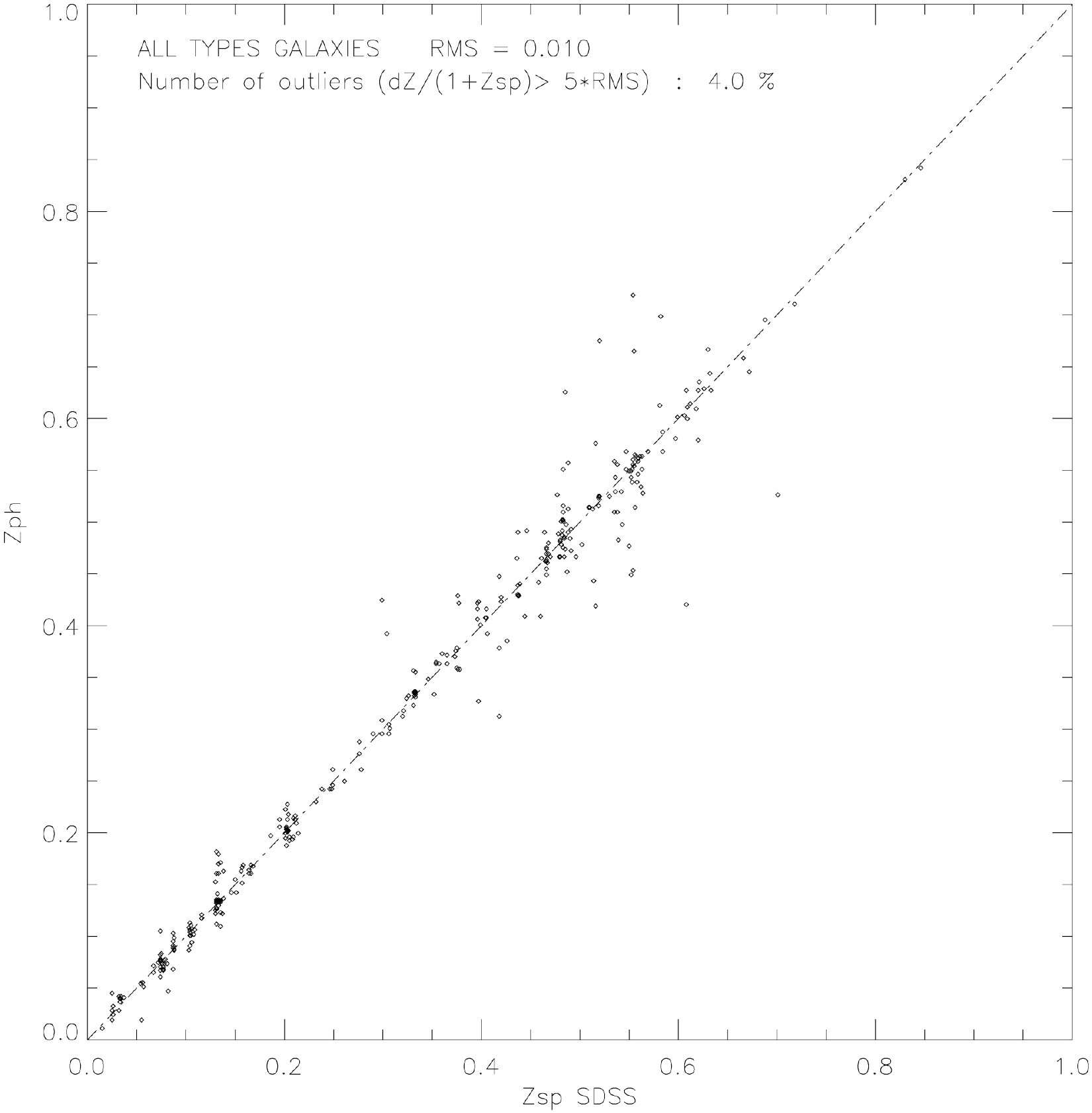}
\caption{Comparison of the photometric redshifts $ z_ \mathrm {ph} $ of the galaxies obtained with the ZEBRA \citep {Feldmann2006} software package with the spectroscopic redshifts $ z_ \mathrm {sp} $ of the galaxies taken from SDSS \citep{Guo2013,Guo2014}  for 473 galaxies with known spectroscopic redshifts. The accuracy of determining the photometric redshift is $ \sigma_z <0.01 $, the outliers percentage is $ \Delta z / (1 + z)> 5. * \sigma_z \sim 4.0 \% $.}
\label{fig_2}
\end{figure}

Using the results of photometry in 17 filters ($ \mathrm {u \_SDSS} $ and 16 mid-band filters) low-resolution spectral energy distributions (SED) were constructed for all galaxies in the sample. The resulting energy distributions were used to estimate the photometric redshift and classification of galaxies.

The method for determining the photometric redshift and SED type of galaxies is based on the correspondence of the spectral pattern of the galaxy to the observed energy distribution. The spectra from \cite {Dodonov2008} and the ZEBRA software package \citep[Zurich's Extragalactic Bayesian Redshift Analyzer,] [] {Feldmann2006} were used as reference spectra for determining photometric redshifts.

The accuracy of determining the photometric redshifts of all types of galaxies was $ \sigma_z <0.01 $, and the percentage of outliers ($ \Delta z / (1 + z)> 5. * \Sigma_z $) $ \sim 4.0 \% $ (Fig. 2). The accuracy of $ \sigma_z $ varies from 0.01 for objects with a magnitude of $ 16 ^ m - 21 ^ m $ in the filter $ \mathrm {r \_SDSS} $ to 0.03 for $ 21 ^ m - 23 ^ m $. Errors in determining the type of galaxy do not exceed $ \sim 3.0 \% $ of the total number of objects at magnitudes weaker than $ R_{AB} = 22 ^ m $. 

\subsection {Galaxy Sample}
A sample of galaxies for study was made from a complete photometric catalog (about 100,000 objects) according to the following criteria:
\begin {enumerate}
\item Objects brighter then $R_{AB}=23$ mag (AB magnitude in R filter)
\item Extended index \citep{Bertin1996} $\textless \; 0.8$ for the objects with $R_{AB}<21$ mag;

Extended index $\textless \; 0.9$ for the objects with $21 \; \mathrm{mag} \; <R_{AB}<22$ mag;

Extended index $\textless \; 0.96$ for the objects with $R_{AB}<23$ mag  
\item Index of contamination $\leq 2$
\end {enumerate} 

Applying these four criteria to the initial sample of more than 100,000 objects, we obtained a sample of 28,398 galaxies that fully satisfies the given conditions. Figure 3 shows the distribution of galaxies depending on the redshift, as well as a comparison with a similar sample of galaxies from the COSMOS survey \cite {Scoville2013}.

\begin{figure}[h]
\hspace*{10pt}
\includegraphics[width=0.9\linewidth]{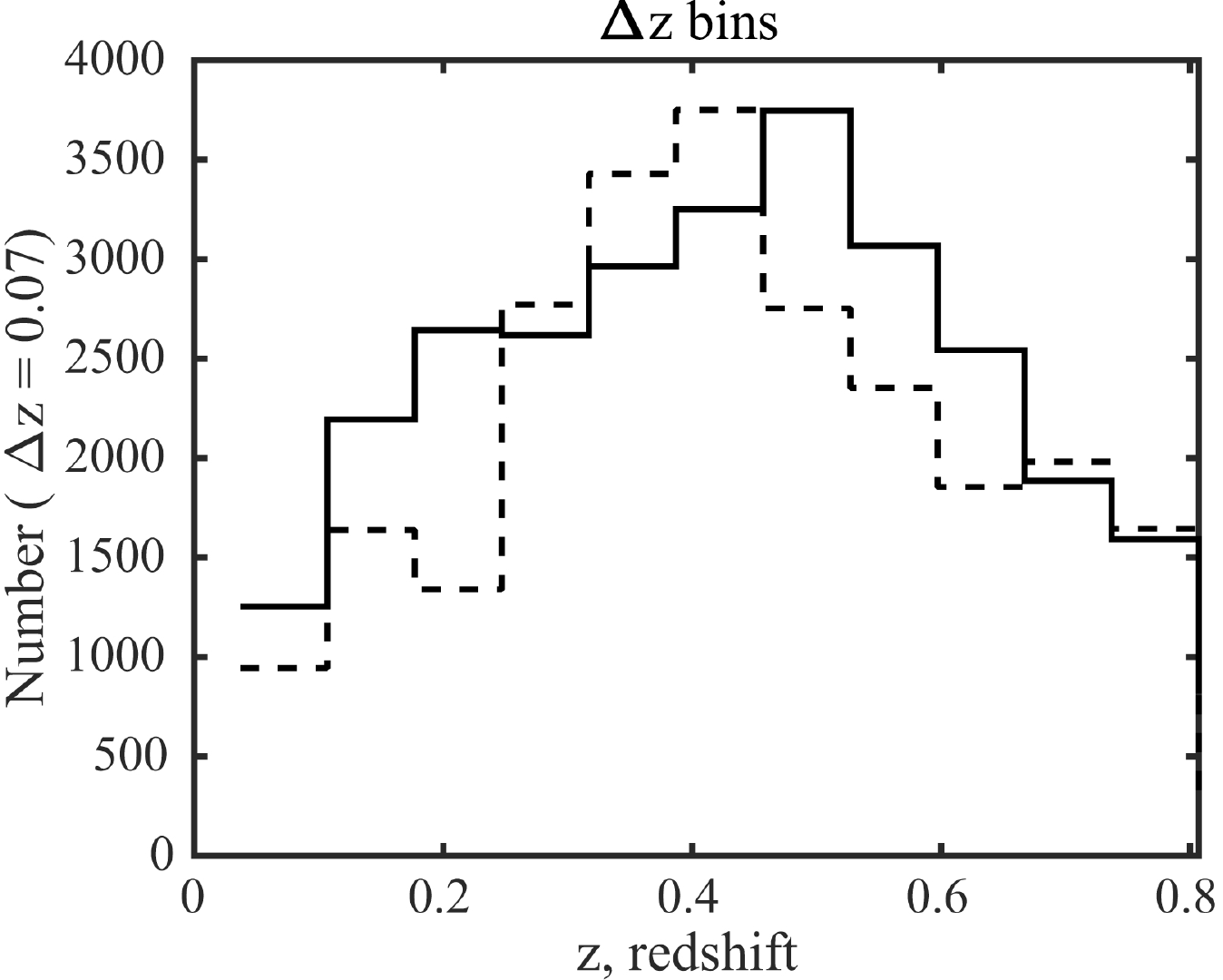}
\caption{The distribution of the number of galaxies as a function of redshift for all types of galaxies for the sample used in this work (solid line) and the sample from COSMOS (dashed line) \cite {Scoville2013} truncated to our selection criteria ($ R_ {AB} <23 $ mag for $ 0 <z \leq 0.8 $). The redshift bins are $ \Delta z = 0.07 $.}
\label{fig_3}
\end{figure}

\subsection {Classification of galaxies}
The spectral types of galaxies were determined using the ZEBRA \cite {Feldmann2006} software package and reference spectra for determining the spectral type of galaxies from \cite {Dodonov2008}. To further study the physical properties of galaxies and their dependencies, the full sample of galaxies was divided into three groups according to the types of galaxies corresponding to the templates:
\begin {enumerate}
\item Early Type Galaxies E - Sa;
\item Late Type Galaxies Sab - Sd;
\item Irregular galaxies and starburst galaxies IRR / SB.
\end {enumerate}

The redshift distribution of galaxies for all field galaxies depending on the spectral type is shown in Fig. 4. This figure also shows data from a similar sample of galaxies from \cite {Ilbert2008}.

The high accuracy of photometric redshifts and deep photometry make it possible to solve the problem of studying the full range of large-scale structures with a large number of objects at large redshifts (up to $ z \sim 0.8 $) and minimizing the effects of fluctuations in the spatial distribution of galaxies due to the sufficient area of the studied field.

\begin{figure}[h]
\hspace*{10pt}
\includegraphics[width=0.9\linewidth]{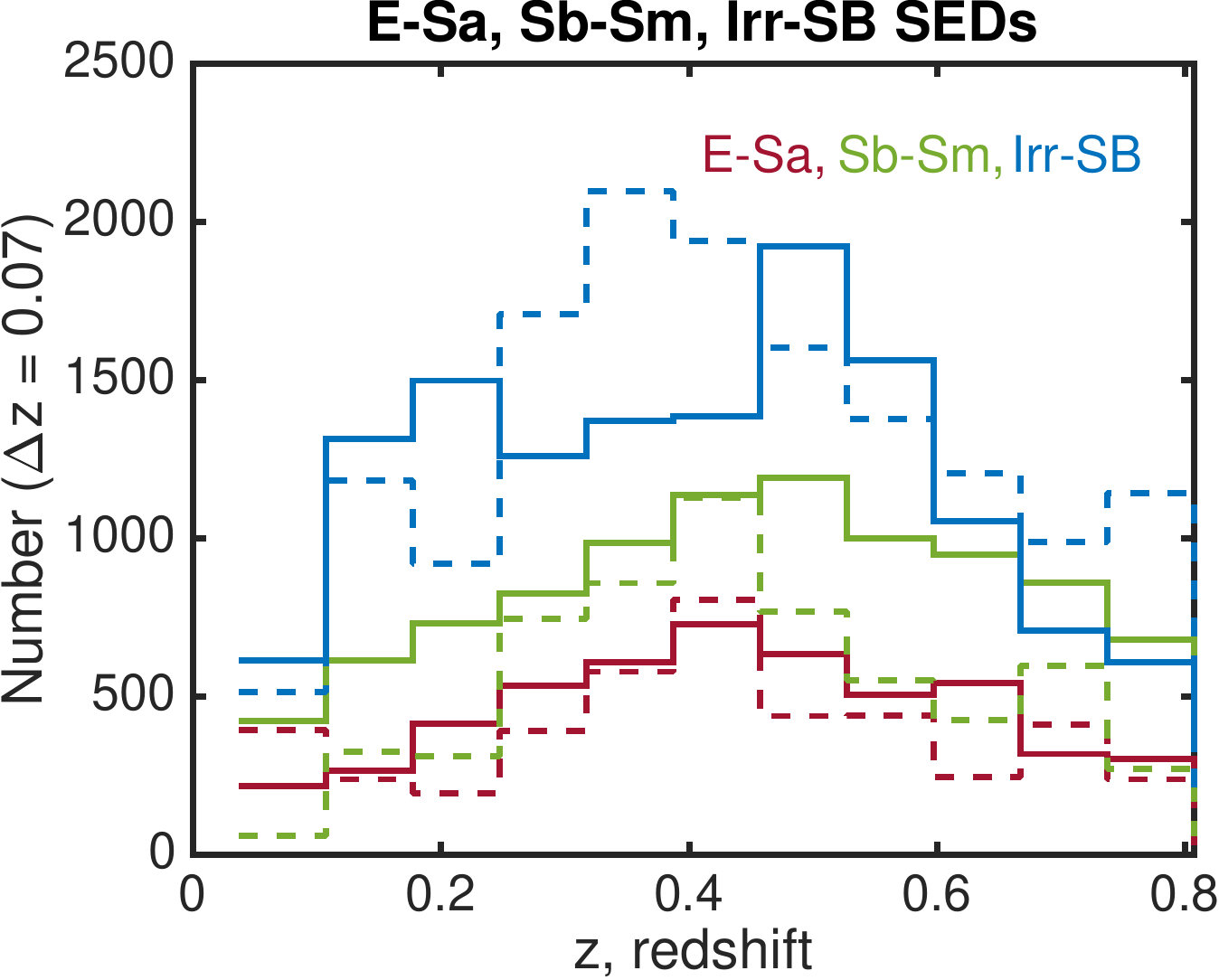}
\caption{The distribution of the number of galaxies as a function of redshift, divided for three groups of galaxies for the sample used in this work (solid lines), and the sample from COSMOS (dashed lines) \cite {Scoville2013} truncated to our selection criteria ($ R_ {AB} <23 $ mag at $ 0 <z \leq 0.8 $). The redshift bins are $ \Delta z = 0.07 $.}
\label{fig_4}
\end{figure}

\section {Determining the density of the environment around galaxies}
The environmental density for each galaxy was obtained from the local surface density of galaxies within narrow redshift layers (Section 4.1) based on photometric redshifts for 28,398 galaxies from the catalog obtained with the 1-meter Schmidt telescope (BAO NAS). Two methods were used to determine overdensities in the large-scale distribution of galaxies: an algorithm with an adaptive aperture and smoothing and two-dimensional Voronoi diagrams \cite {Grokh2019}.

\subsection {Narrow redshift layers}
To determine the overdensities in the galaxy distribution, the consistency between the width of the layer in which the structures are separated and the accuracy of determining the photometric redshift is important. Using too thin layers by redshift leads to an undetermination of structures and their members. Wide layers are characterized by the projection of galaxies with a different redshift onto large-scale structures to which they do not belong.

The accuracy of determining photometric redshifts for the galaxy catalog is $ \sigma_z <0.01 $. Based on this, the redshift layer width is $ \Delta z = 2 * \sigma_z * (1 + z) $ or $ \Delta z = 0.02 * (1 + z) $. To each interval $ 25 \% $ of its value was added on each side to avoid losses in the determination of large-scale structures at the layer boundary.

\subsection {Structure determination methods}
We used two methods for determining overdensities in the distribution of galaxies: an algorithm with an adaptive aperture and smoothing of the density of the environment and two-dimensional Voronoi tessellation \cite {Grokh2019}.

\begin{figure}[h]
\hspace*{10pt}
\includegraphics[width=1.0\linewidth]{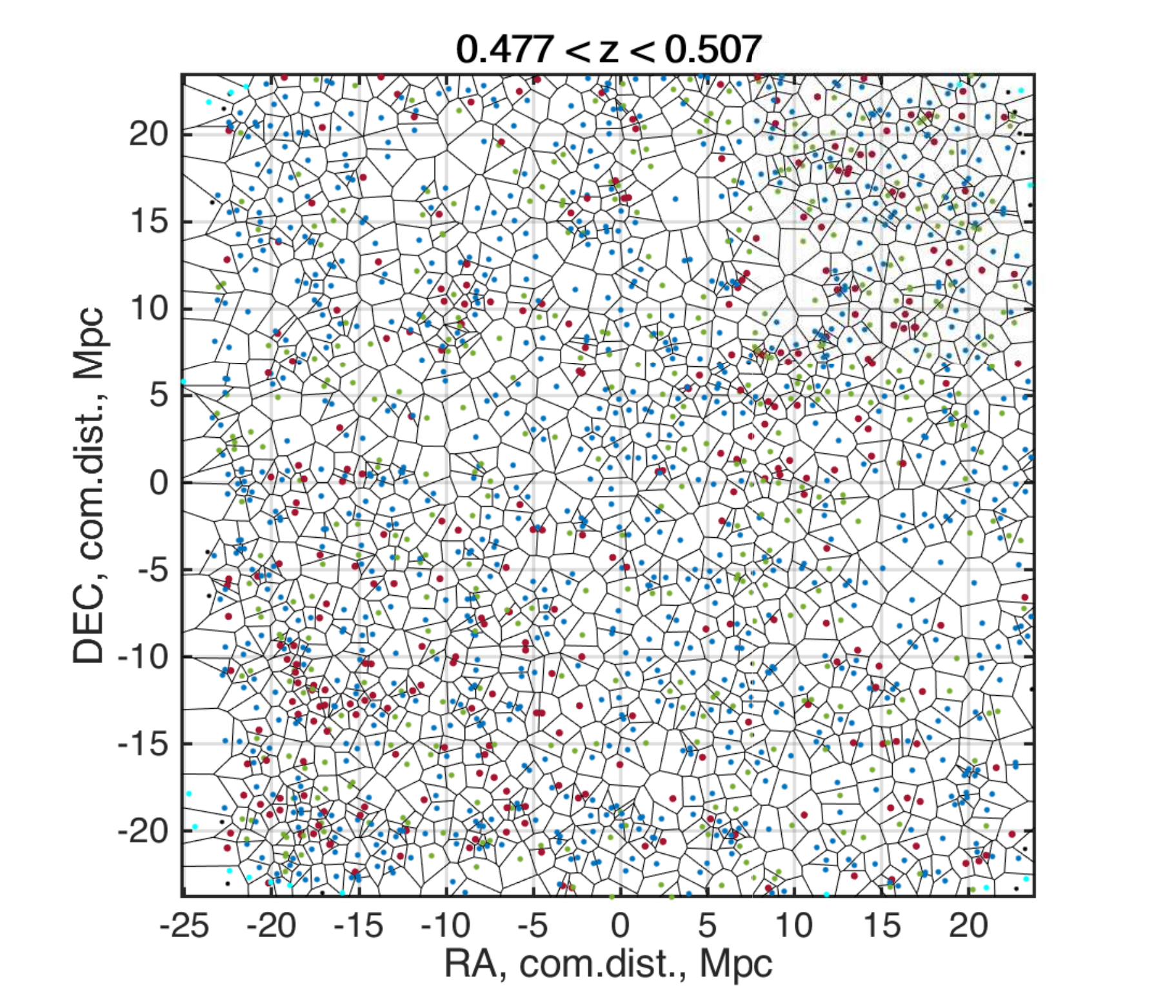}
\caption{Voronoi tessellation constructed for the $ 0.477 \leq z \leq 0.507 $ layer. Voronoi regions in two-dimensional space show regions, each point of which is closer to an individual galaxy from the sample than to any other. On the diagram, individual galaxies are shown by red, green, or blue dots depending on the type of SED galaxy (galaxies of early types E - Sa, late types Sab - Sd and irregular galaxies  IRR / starburst galaxies SB, respectively). Light blue dots indicate galaxies for which Voronoi regions are not closed; such galaxies are excluded from the density calculation in the future.}
\label{fig_5}
\end{figure}

For Voronoi tessellation, each galaxy is at a photometric redshift defined by the ZEBRA \citep {Feldmann2006} software package in Maximum Likelihood mode (maximum likelihood method). Since we use two-dimensional Voronoi tessellation, a galaxy is considered to belong to a particular layer if its redshift (without taking into account uncertainty) falls into the redshift interval of the layer. Since the edge galaxies do not have Voronoi closed regions (Fig. 5), the number of galaxies in the sample for Voronoi tessellation is less than 28,398 galaxies of the full sample and is 27,446.

For the Voronoi algorithm, a step was added with interpolation of the values of the density of the environment obtained in accordance with \cite {Grokh2019} (Fig. 6).

\begin{figure*}[h!]
\begin{minipage}[h]{0.4\linewidth} 
\center{\includegraphics[scale=0.4]{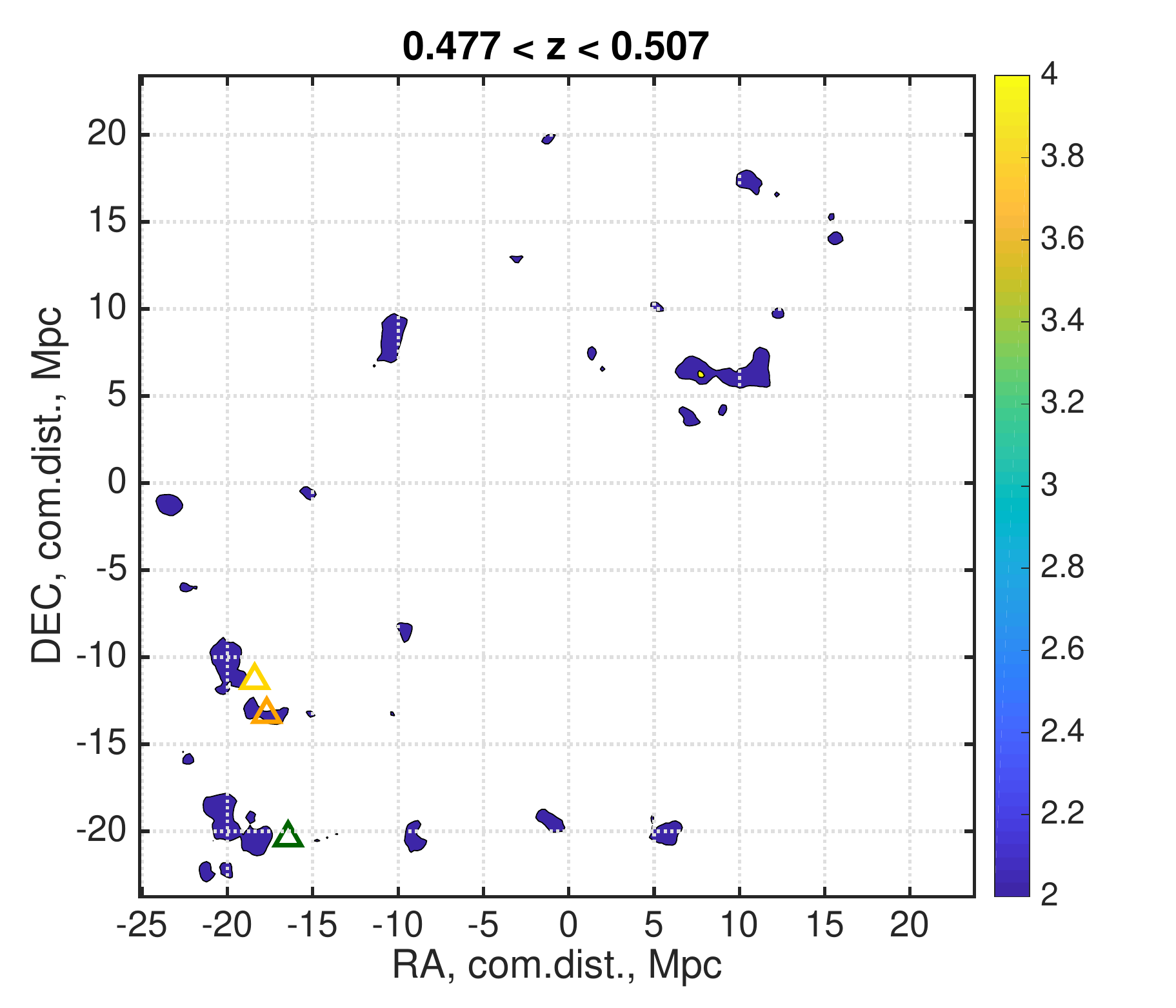}}\\ \end{minipage}
\hfill
\begin{minipage}[h]{0.4\linewidth}  
\center{ \includegraphics[scale=0.4]{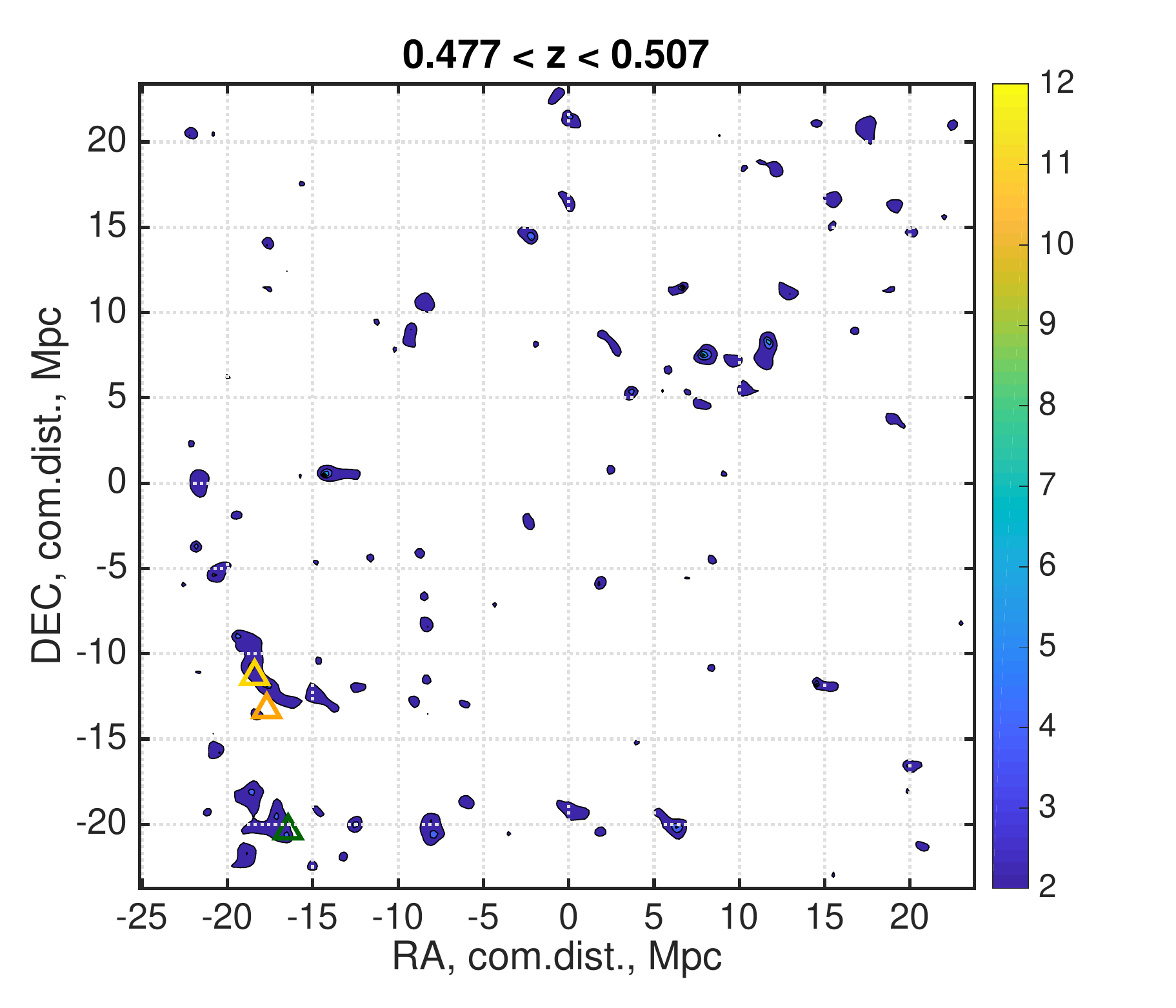}}\\ \end{minipage}
\caption{Density contrast maps constructed for the $ 0.477 \leq z \leq 0.507 $ layer by the adaptive aperture method with smoothing the density of the environment (left) and the Voronoi tessellation (right). The color bar shows how many times the density of the environment is exceeded in comparison with the average density in the layer. Triangles indicate the centers of clusters WHL J094637.9 + 471440 (yellow), WHL J094645.7 + 471107 (orange), WHL J094659.9 + 465810 (green) from the \cite {Wen2015} catalog, which were detected by radiation in the X-ray range waves. These clusters are also detected by both of our methods. A complete set of 57 redshift slices is available on \href{https://github.com/ale-gro/density_maps}{https://github.com/ale-gro/densitymaps}.}
\label{fig_6}
\end{figure*}

To work with the observational data of the 1-meter Schmidt telescope, an algorithm with an adaptive aperture and smoothing the contrast of the density of the environment for each of the studied galaxies was also used. The need for smoothing by redshift is due to the fact that the photometric redshift is not determined exactly, but has a certain error. Therefore, when isolating large-scale structures, the density of the environment was first calculated using the formulas described in \cite {Grokh2019}. Then, the density of the environment for each galaxy was smoothed by a Gaussian filter at redshift with a filter width of $ 2 \sigma_z $. Thus, a pseudo-three-dimensional distribution of the density of the environment was obtained for each galaxy, taking into account the uncertainty of its redshift in a thin spatial section. Next, three-dimensional interpolation of the obtained values and projection of the obtained three-dimensional results on a two-dimensional picture plane (Ra-Dec plane, Fig. 6) were performed.

\begin{figure}[h]
\hspace*{10pt}
\includegraphics[width=1.0\linewidth]{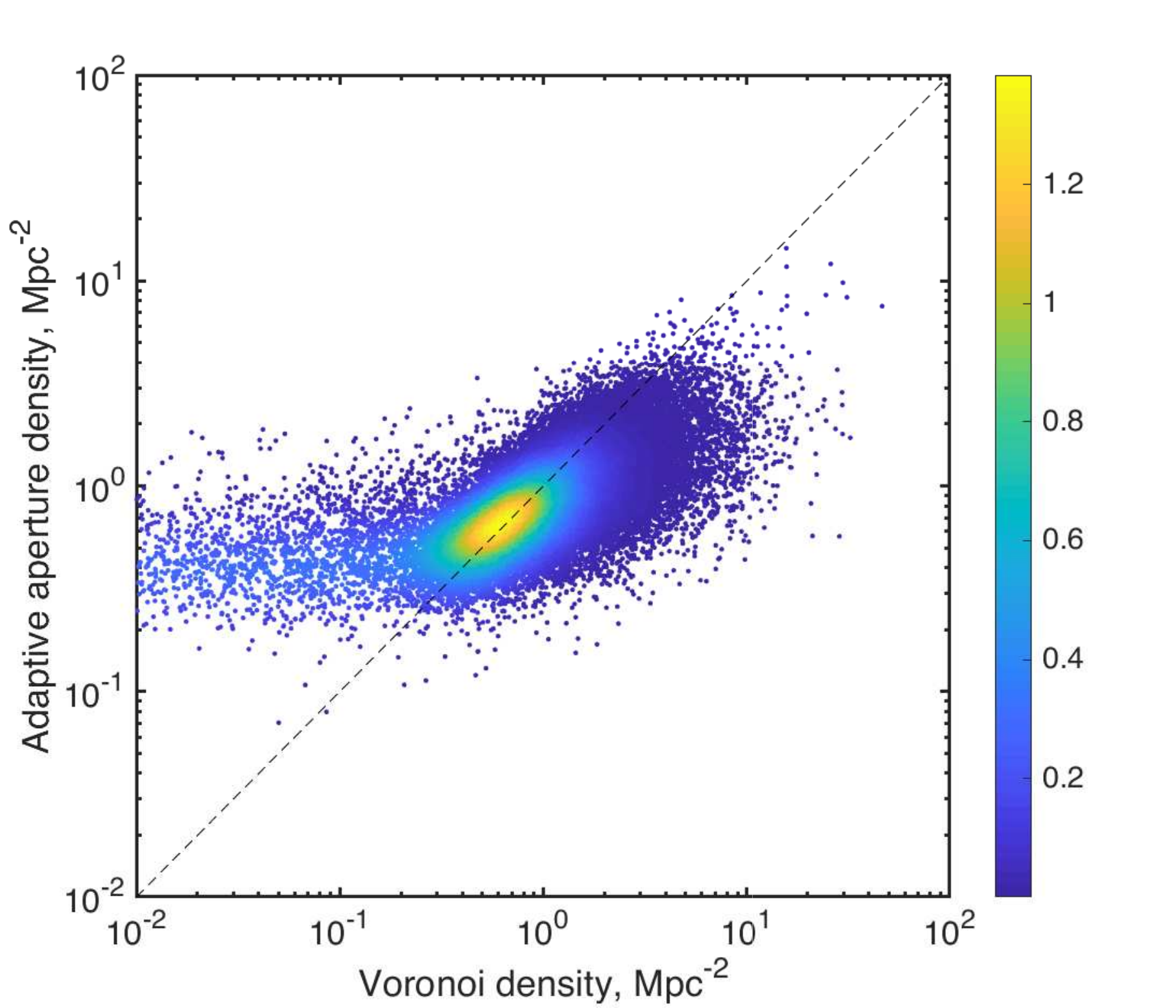}
\caption{Comparison of the surface density of the environment for each galaxy obtained using the Voronoi algorithm and the density obtained by the algorithm with an adaptive aperture and smoothing the density of the environment. The color indicates the probability density for each galaxy. Over three orders of magnitude of the densities, both algorithms give a similar result for a large number of objects, close to the straight line 1: 1, indicated by the dashed line. The outlier from this distribution, which is in the leftmost position, can be explained by the fact that the adaptive aperture algorithm extracts only significant overdensities, while the Voronoi algorithm determines any excess density over the average value.}
\label{fig_7}
\end{figure}

Each of the methods (Voronoi tessellation, an algorithm with adaptive aperture and smoothing) has its advantages and disadvantages. Moreover, the methods give a consistent result in the determination of large-scale structures in the distribution of galaxies, which provides confidence in the results obtained (Fig. 7). So, using both methods, overdensities were found that correspond to almost all known clusters and groups of galaxies in the field HS47.5-22. A complete list of 44 field clusters is given in Table. \ref {tabl2}, of which 42 clusters were detected. All reconstructed clusters and groups of galaxies are found within the error of determining their redshift $ \delta_z $, which is defined as half the width of a narrow redshift layer (Fig. 9). We have not detected two clusters of galaxies: PDCS 043 ($ \mathrm {z_ {phot}} = 0.200 $) and WHL J094840.1 + 475045 ($ \mathrm {z_ {spec}} = 0.393 $). In the comments, the authors of {\cite {Postman1996}} write that PDCS 043 was not found in the filter I and has a small number of cluster members. In the center of the WHL J094840.1 + 475045 cluster, there are two galaxies with $ \mathrm {z_ {spec}} = 0.396 $ and $ \mathrm {z_ {spec}} = $ 0.399, and only two galaxies with $ \mathrm {z_ {phot}} = 0.393 \pm 0.015 $ are found in 1 Mpc, (Fig. 8). Apparently, the cluster of galaxies identified by the authors {\cite {Wen2015}} is erroneous.

\begin{figure}[h]
\hspace*{10pt}
\includegraphics[width=1.0\linewidth]{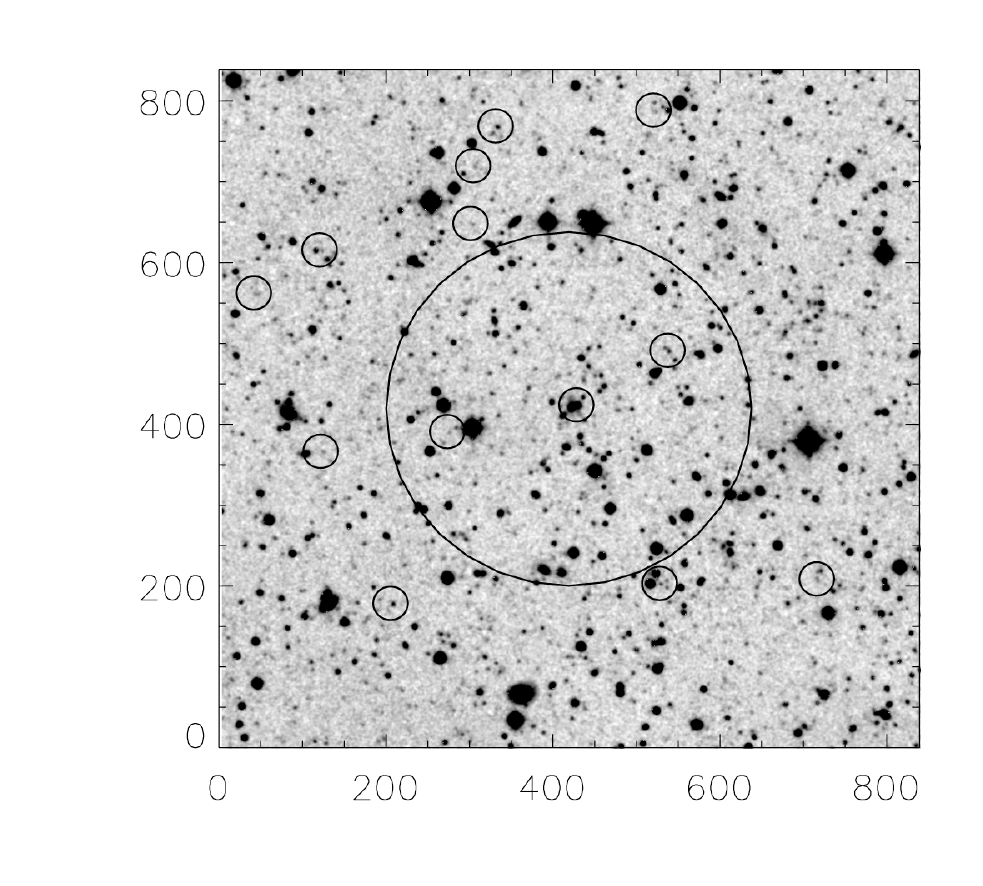}
\caption{The map the WHL J094840.1 + 475045 cluster environment (15 * 15 arcmin). Small circles show galaxies with $ \mathrm {z_ {phot}} = 0.393 \pm 0.015 $, the radius of a circle of large diameter is 1 Mpc.}
\label{fig_8}
\end{figure}

\begin{table*}[tp!]
\caption{Known clusters of the HS47.5-22 field detected by the methods of Voronoi diagrams and an algorithm with adaptive aperture and smoothing from \cite{Postman1996}, \cite{Koester2007}, \cite{Wen2015}, \cite{Wen2010}, \cite{Holden1999}, \cite{McConnachie2009}, \cite{Hao2010}, \cite{Smith2012}.} 
\centering 
\begin{tabular}{c c c c c c c} 
\hline\hline 
\# & Name & Coordinates & Coordinates & z & $\mathrm{z_{phot}}$ & $\delta_z$\\
  & & of the center, RA & of the center, DEC & (NED\footnote{The NASA/IPAC Extragalactic Database (NED) is operated by the Jet Propulsion Laboratory, California Institute of Technology, under contract with the National Aeronautics and Space Administration.}) &  & \\[0.5ex]
\hline 
1 & PDCS $043$ & $09^h 52^m 15.1^s$ & $+47^d 57^m 44^s$ & 0.2000 & - & -\\
2 & PDCS $040 $ & $09^h 53^m 25.6^s$ & $+47^d 58^m 55^s$ & 0.2028 & 0.205 & 0.011 \\
3 & PDCS $042 $ & $09^h 53^m 54.3^s$ & $+48^d 00^m 04^s$ & 0.1992 & 0.199 & 0.012\\
4 & PDCS $039 $ & $09^h 51^m 25.2^s$ & $+47^d 49^m 50^s$ & 0.6000 & 0.484 & 0.014\\
5 & PDCS $008s $ & $09^h 53^m 49.9^s$ & $+47^d 52^m 28^s$ & 0.6000 & 0.698 & 0.017 \\
6 & PDCS $037 $ & $09^h 51^m 41.5^s$ & $+47^d 41^m 30^s$ & 0.6000 & 0.648 & 0.017 \\
7 & PDCS $035 $ & $09^h 52^m 31.2^s$ & $+47^d 36^m 27^s$ & 0.6000 & 0.648 & 0.017\\
8 & PDCS $036 $ & $09^h 53^m 53.7^s$ & $+47^d 40^m 15^s$ & 0.2480 & 0.247 & 0.013\\
9 & PDCS $033 $ & $09^h 52^m 13.1^s$ & $+47^d 16^m 48^s$ & 0.6400 & 0.600 & 0.016\\
10 & PDCS $006s $ & $09^h 49^m 56.5^s$ & $+47^d 06^m 40^s$ & 0.4000 & 0.462 & 0.015\\
11 & PDCS $029 $ & $09^h 53^m 12.1^s$ & $+47^d 08^m 58^s$ & 0.4000 & 0.412 & 0.014\\
12 & WHL J$095357.5$ & $09^h 53^m 57.5^s$ & $+48^d 14^m 31^s$ & 0.13505 & 0.135 & 0.011\\
 & $+481431 $ &  &  &  & &\\
13 & WHL J$095323.9$ & $09^h 53^m 23.92^s$ & $+47^d 58^m 42.5^s$ & 0.2030 & 0.204 & 0.011\\
 & $+475842 $ &  &  & & & \\
14 & WHL J$095312.2$ & $09^h 53^m 12.17^s$ & $+47^d 38^m 22.3^s$ & 0.5543 & 0.568 & 0.016\\
 & $+473822 $ &  &  &  & &\\ 
15 & WHL J$095027.9$ & $09^h 50^m 27.86^s$ & $+48^d 14^m 35.3^s$ & 0.2085 & 0.199 & 0.012\\
 & $+481435 $ &  &  &  & &\\ 
16 & WHL J$094930.4$ & $09^h 49^m 30.4^s$ & $+48^d 17^m 56^s$ & 0.2047 & 0.199 & 0.012\\
 & $+481756 $ &  &  &  & &\\ 
17 & WHL J$094933.0$ & $09^h 49^m 32.99^s$ & $+47^d 40^m 49.7^s$ & 0.5554 & 0.552 & 0.016 \\
 & $+474050 $ &  &  & & & \\ 
18 & WHL J$094840.1$ & $09^h 48^m 40.1^s$ & $+47^d 50^m 45^s$ & 0.3934 & - & - \\
 & $+475045 $ &  &  & & & \\ 
19 & WHL J$094657.3$ & $09^h 46^m 57.3^s$ & $+48^d 15^m 26.1^s$ & 0.3745 & 0.378 & 0.014\\
 & $+481526 $ &  &  & & & \\
20 & WHL J$094557.8$ & $09^h 45^m 57.78^s$ & $+48^d 10^m 35.2^s$ & 0.3451 &  0.350 & 0.013 \\
 & $+481035 $ &  &  &  & &\\
21 & WHL J$094651$ & $09^h 46^m 51.8^s$ & $+48^d 18^m 16^s$ & 0.3521 &  0.364 & 0.014 \\
 & $+481816 $ &  &  &  &  &\\ 
 22 & WHL J$094913.0$ & $09^h 49^m 12.99^s$ & $+47^d 22^m 47.7^s$ & 0.3061 &  0.310 & 0.013 \\
 & $+472248 $ &  &  &  &  &\\ 
23 & WHL J$094725.0$ & $09^h 47^m 24.97^s$ & $+47^d 17^m 46.6^s$ & 0.3249 & 0.324 & 0.013\\
 & $+471747 $ &  &  &  &  &\\ 
24 & WHL J$094637.9$ & $09^h 46^m 37.9^s$ & $+47^d 14^m 40.2^s$ & 0.4830 & 0.484 & 0.015\\
 & $+471440 $ &  &  &  &  &\\ 
25 & WHL J$094645.7$ & $09^h 46^m 45.74^s$ & $+47^d 11^m 07.4^s$ & 0.4805 & 0.492 & 0.015\\
 & $+471107 $  &  &  &  &  &\\ 
26 & WHL J$094659.9$ & $09^h 46^m 59.92^s$ & $+46^d 58^m 10.2^s$ & 0.4878 & 0.492 & 0.015\\
 & $+465810 $  &  &  &  &  &\\ 
27 & WHL J$094642.2$ & $09^h 46^m 42.2^s$ & $+46^d 58^m 57.0^s$ & 0.5323 & 0.537 & 0.015\\
 & $+465857 $  &  &  &  &  &\\ 
28 & WHL J$094827.1$ & $09^h 48^m 27.1^s$ & $+46^d 59^m 35.1^s$ & 0.3960 & 0.405 & 0.014 \\
 & $+465935 $  &  &  &  &  &\\ 
29 & WHL J$094615.2$ & $09^h 46^m 15.2^s$ & $+47^d 00^m 16^s$ & 0.3030 & 0.298 & 0.013\\
 & $+470016 $  &  &  &  &  &\\ 
30 & WHL J$094957.2$ & $09^h 49^m 57.2^s$ & $+47^d 10^m 31^s$ & 0.4050 & 0.412 & 0.014 \\
 & $+471031 $  &  &  &  &  &\\  
31 & WHL J$094952.4$ & $09^h 49^m 52.3^s$ & $+46^d 59^m 34^s$ & 0.3520 & 0.364 & 0.014 \\
 & $+465934 $  &  &  &  &  &\\ 
32 & WHL J$095328.5$ & $09^h 53^m 28.53^s$ & $+46^d 57^m 07.7^s$ & 0.2479 & 0.247 & 0.013\\
 & $+465708 $  &  &  &  &  &\\  
[1ex] 
\hline 
\end{tabular}
\label{tabl2} 
\end{table*}

\begin{table*}[tp!]
\tablename{ \textbf{ Table 2.} Continuation.} 
\centering 
\begin{tabular}{c c c c c c c} 
\hline\hline 
\# & Name & Coordinates & Coordinates & z & $\mathrm{z_{phot}}$ & $\delta_z$\\
  & & of the center, RA & of the center, DEC & (NED) & &  \\[0.5ex]
\hline 
33 & GHO $0947+4758 $ & $09^h 51^m 09.90^s$ & $+47^d 43^m 54.0^s$ & 0.3342 & 0.344 & 0.014 \\
34 & GHO $0949+4732 $ & $09^h 52^m 29.00^s$ & $+47^d 17^m 49.0^s$ & 0.3000 & 0.235 & 0.013 \\
35 & SDSSCGB $00840 $ & $09^h 51^m 18.60^s$ & $+48^d 13^m 18.0^s$ & 0.1320 & 0.129 & 0.011\\
36 & GMBCG J$148.37587$ & $09^h 53^m 30.20^s$ & $+47^d 40^m 51.0^s$ & 0.3770 & 0.364 & 0.014 \\
 & $+47.68093 $   &  &  &  &  &\\ 
37 & NSC J$095412$ & $09^h 54^m 12.0^s$ & $+48^d 13^m 07.0^s$ & 0.1933 & 0.141 & 0.012\\
 & $+481348 $   &  &  &  &  &\\ 
38 & SDSSCGB $27386 $ & $09^h 49^m 04.90^s$ & $+47^d 47^m 13.0^s$ & 0.2150 & 0.135 & 0.011 \\
39 & GMBCG J$147.17119$ & $09^h 48^m 41.10^s$ & $+47^d 47^m 48.0^s$ & 0.3620 & 0.364 & 0.016 \\
 & $+47.79663 $   &  &  &  &  &\\
40 & GMBCG J$146.74573$ & $09^h 46^m 59.00^s$ & $+48^d 10^m 09.0^s$ & 0.3170 &  0.311 & 0.013 \\
 & $+48.16905 $  &  &  &\\
41 & GMBCG J$147.32896$ & $09^h 49^m 18.90^s$ & $+47^d 09^m 45.0^s$ & 0.4890 & 0.537 & 0.015 \\
 & $+47.16263 $  &  &  &  &  &\\
 42 & SDSSCGB $04432 $ & $09^h 46^m 45.1^s$ & $+47^d 12^m 58.0^s$ & 0.1040 & 0.096 & 0.011\\
43 & Mr20:[BFW2006] & $09^h 47^m 49.70^s$ & $+47^d 03^m 53.0^s$ & 0.0857 & 0.096 & 0.011\\
 & $22757 $ &  &  &  &  &\\
44 & MSPM $01061 $ & $09^h 49^m 10.0^s$ & $+46^d 58.6^m$ & 0.03282 & 0.038 & 0.010\\
 [1ex] 
\hline 
\end{tabular}
\label{} 
\end{table*}

\begin{figure}[h]
\hspace*{10pt}
\includegraphics[width=1.0\linewidth]{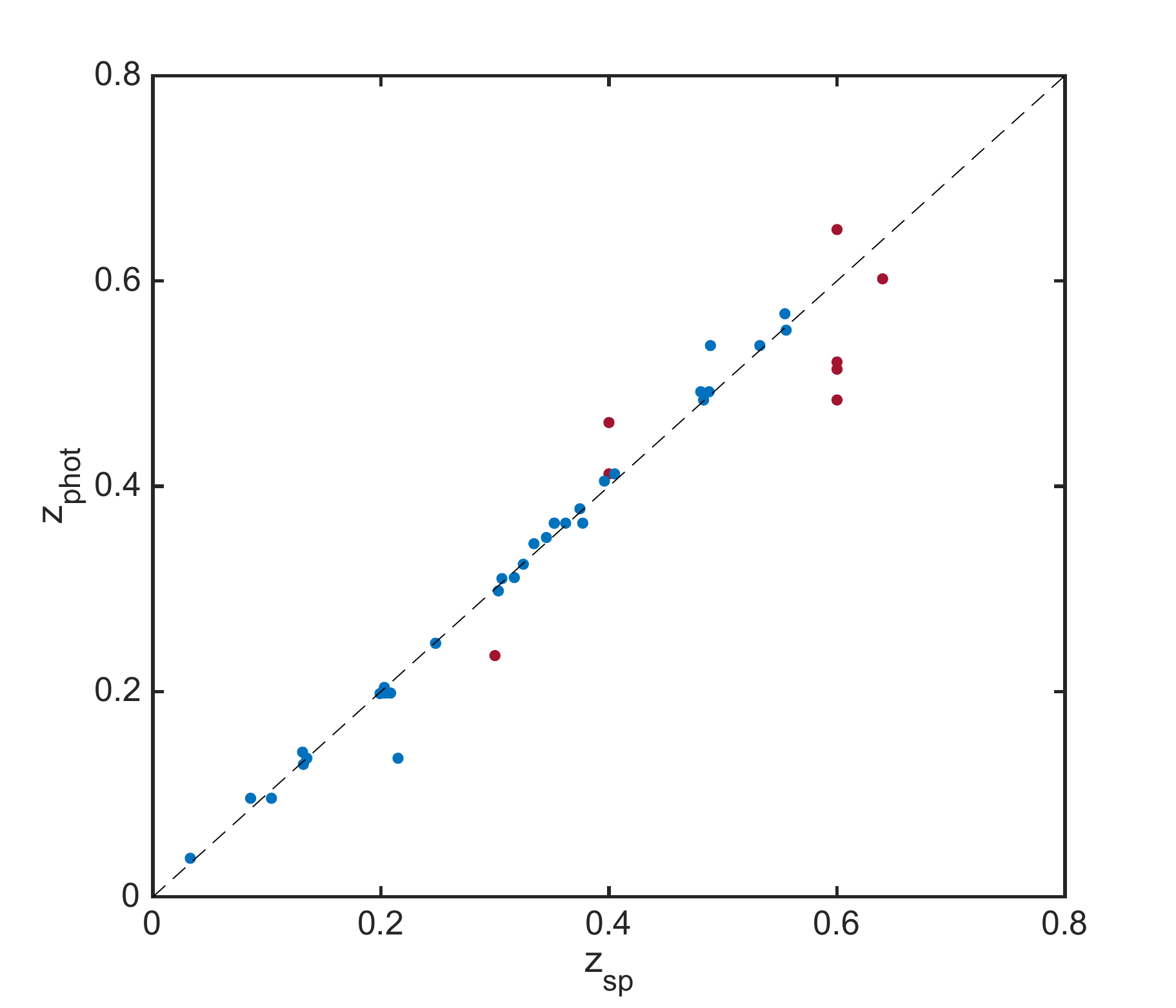}
\caption{Comparison of the photometric redshifts $ z_ \mathrm {ph} $ of the galaxy clusters obtained using the Voronoi diagram methods and the adaptive aperture and smoothing algorithm with the spectroscopic redshifts $ z_ \mathrm {sp} $ of the galaxies taken from \citep {Postman1996, Koester2007, Wen2015, Wen2010, Holden1999, McConnachie2009, Hao2010, Smith2012} for 42 galaxy clusters with known spectroscopic (blue dots) and photometric (red dots) redshifts. The scatter of values in determining the photometric redshift of clusters is $ \delta_z <0.017 $.}
\label{fig_9}
\end{figure}

Both methods determine the two-dimensional surface density of galaxies in each redshift layer and not the true volume density of galaxies. The direct determination of volume densities in three-dimensional mode requires more accurate redshifts. To conduct such studies, the accuracy of the determination of redshifts should be at least 10 times higher, which corresponds to the accuracy of the determination of spectral redshifts.

On the whole, one can expect proportionality between the projected two-dimensional and true three-dimensional densities if the thickness of the layers by redshift is optimally determined (see Section 4.1).

\subsection {Comparison of the algorithm with adaptive aperture and Voronoi tessellation}
Despite the fact that the approaches and limitations of the method with an adaptive aperture and smoothing the density of the environment and Voronoi tessellation to construct density contrast maps are very different, they show a consistent result. In Fig. 6 we shows the distribution of the contrast ratio of the density of the environment determined for each galaxy by the method with adaptive aperture and smoothing and by the method of Voronoi diagrams for a sample of 27,446 galaxies. This number of galaxies is slightly less than the total number of galaxies in the sample due to the fact that the outer Voronoi cells are not closed, and an estimate of the cell area and the density of the environment has not been obtained for these galaxies. Over three orders of magnitude of the densities, both algorithms give a similar result for a large number of objects, close to the straight line 1: 1, indicated by the dashed line.

The results obtained using the method with an adaptive aperture and smoothing of the density of the environment have a higher sample purity (see \cite {Grokh2019}) and reveal only significant large-scale overdensities. The method of Voronoi tessellation restores a more noisy reconstructed sample of overdensities, while the detail of density contrast maps obtained by the method is higher than for maps obtained by the adaptive aperture with smoothing method.

In general, the use of two independent methods for constructing density contrast maps provides confidence in the results obtained. Density contrast maps for the HS 47.5 - 22 field for the entire redshift slice range up to $ z \sim 0.8 $ are available on \href{https://github.com/ale-gro/density_maps}{https://github.com/ale-gro/densitymaps}.

\section {Large-scale distribution structure HS 47.5 - 22}

\begin{figure*}[]
\begin{minipage}[h]{0.45\linewidth} 
\center{\includegraphics[scale=0.45]{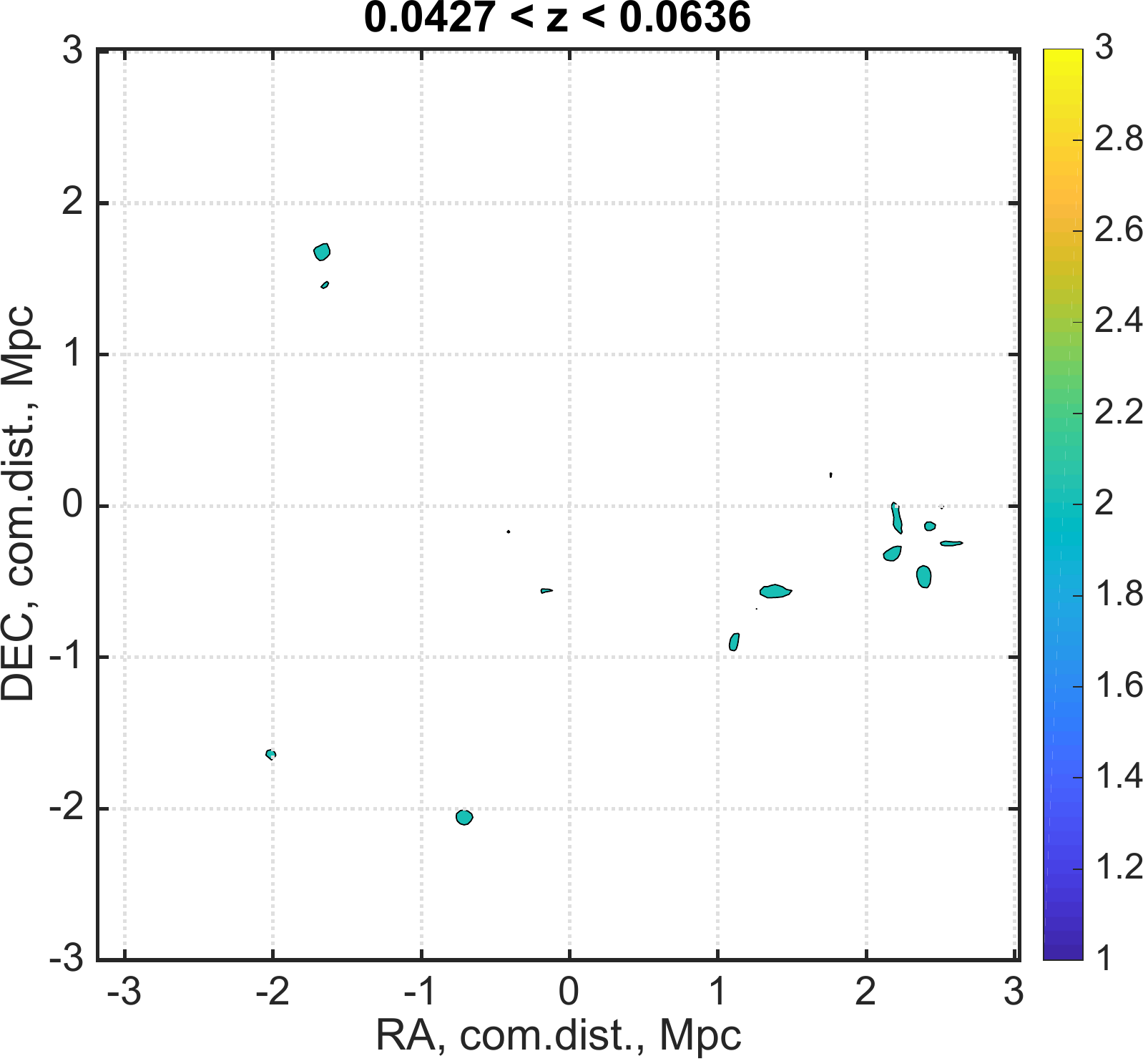}}  \\  
 \end{minipage}
 \hfill
\begin{minipage}[h]{0.45\linewidth}  
\center{\includegraphics[scale=0.45]{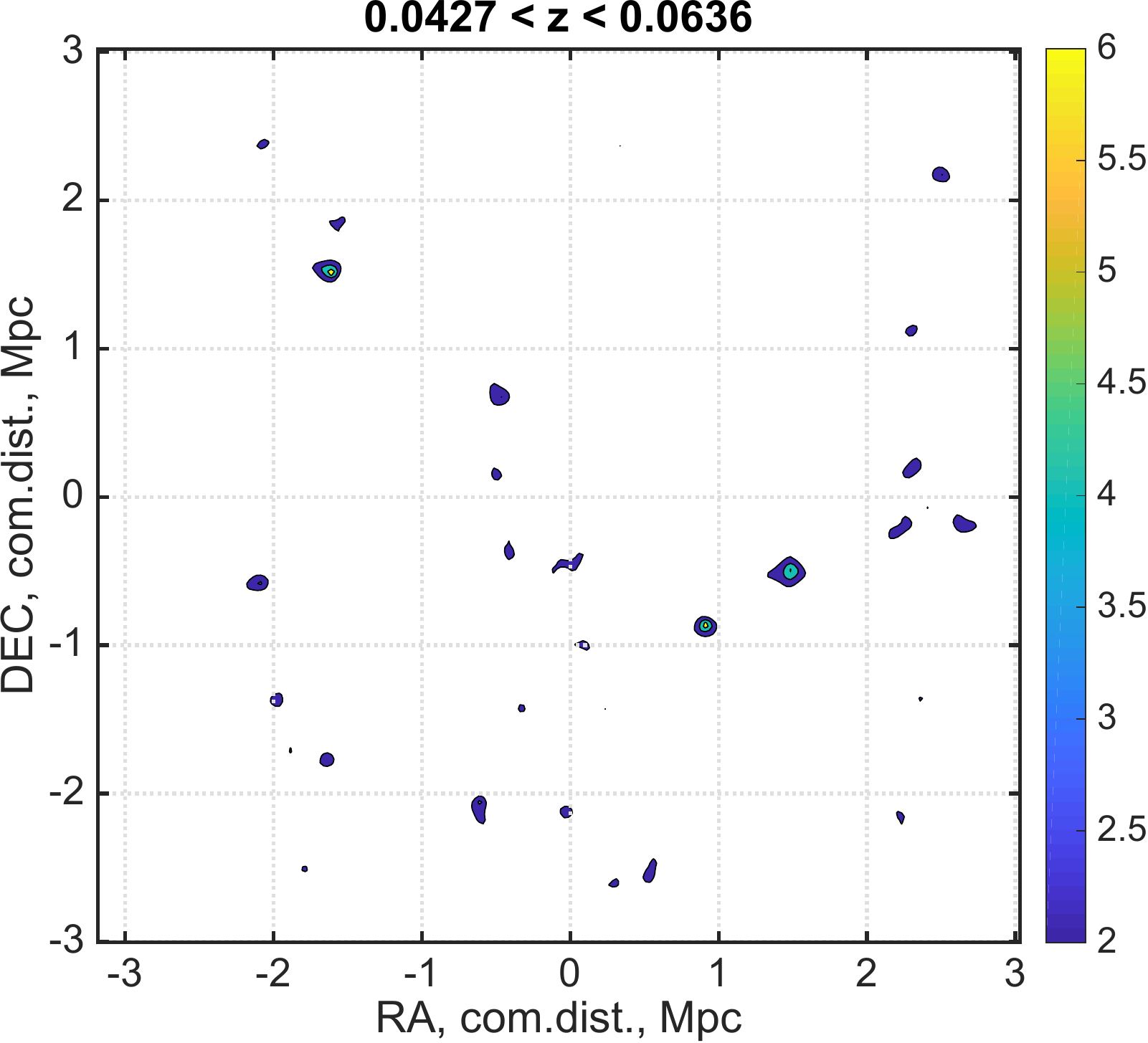}}  \\  
 \end{minipage}
 \vfill
 \begin{minipage}[h]{0.45\linewidth}  
\center{ \includegraphics[scale=0.45]{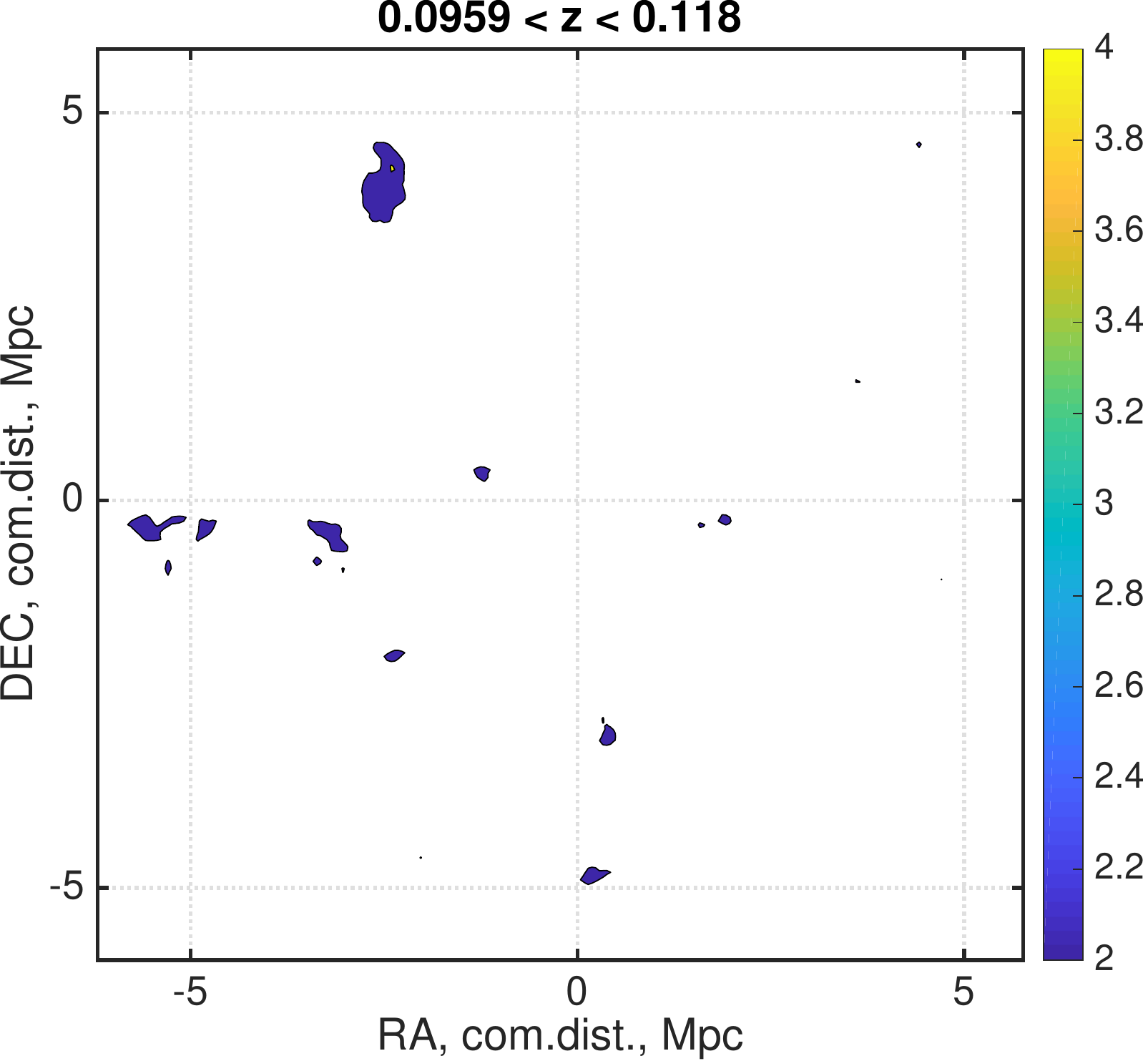}} \\  
 \end{minipage}
  \hfill
\begin{minipage}[h]{0.45\linewidth} 
\center{\includegraphics[scale=0.45]{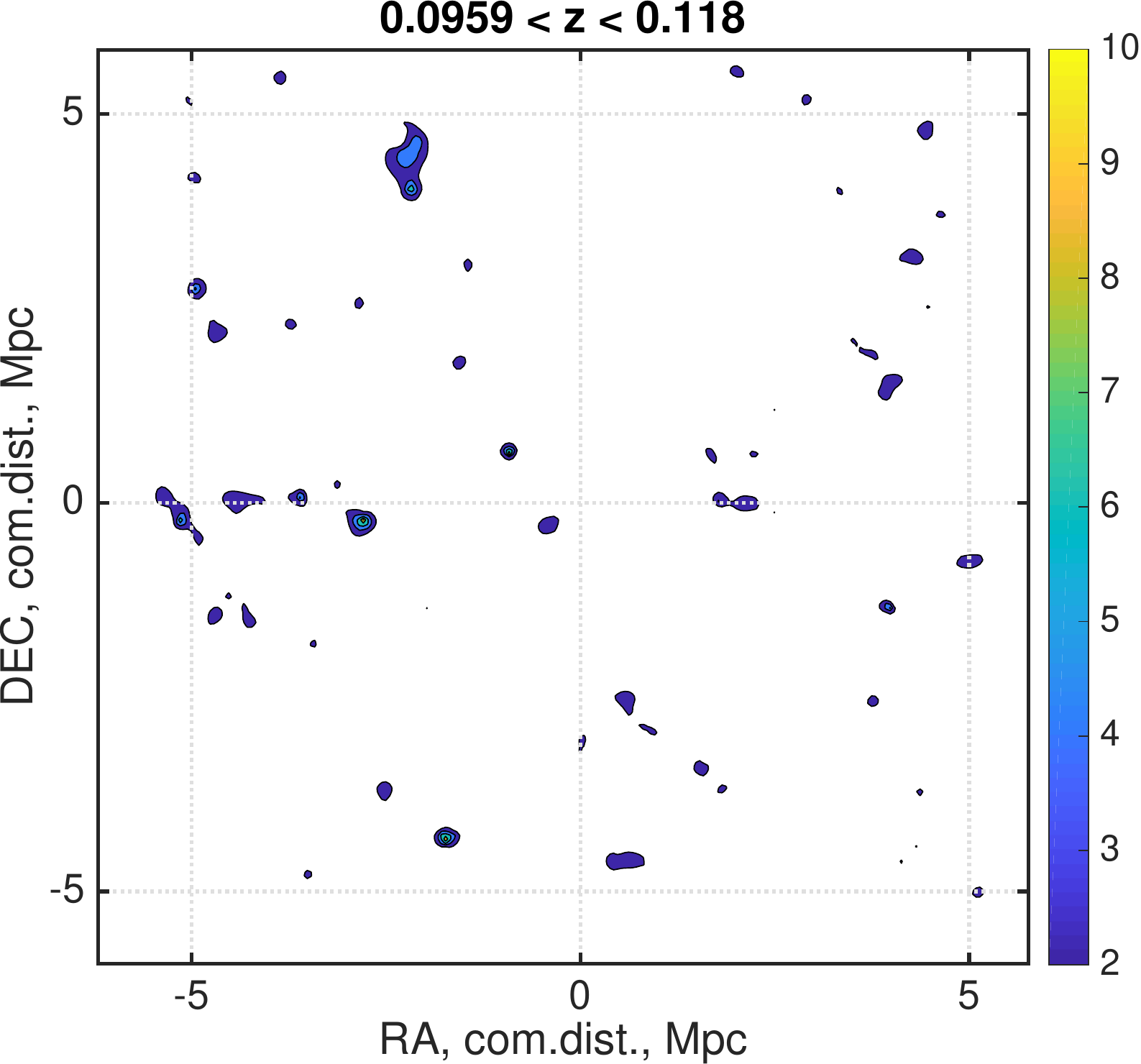}} \\  
 \end{minipage}
  \vfill
 \begin{minipage}[h]{0.45\linewidth}  
\center{ \includegraphics[scale=0.45]{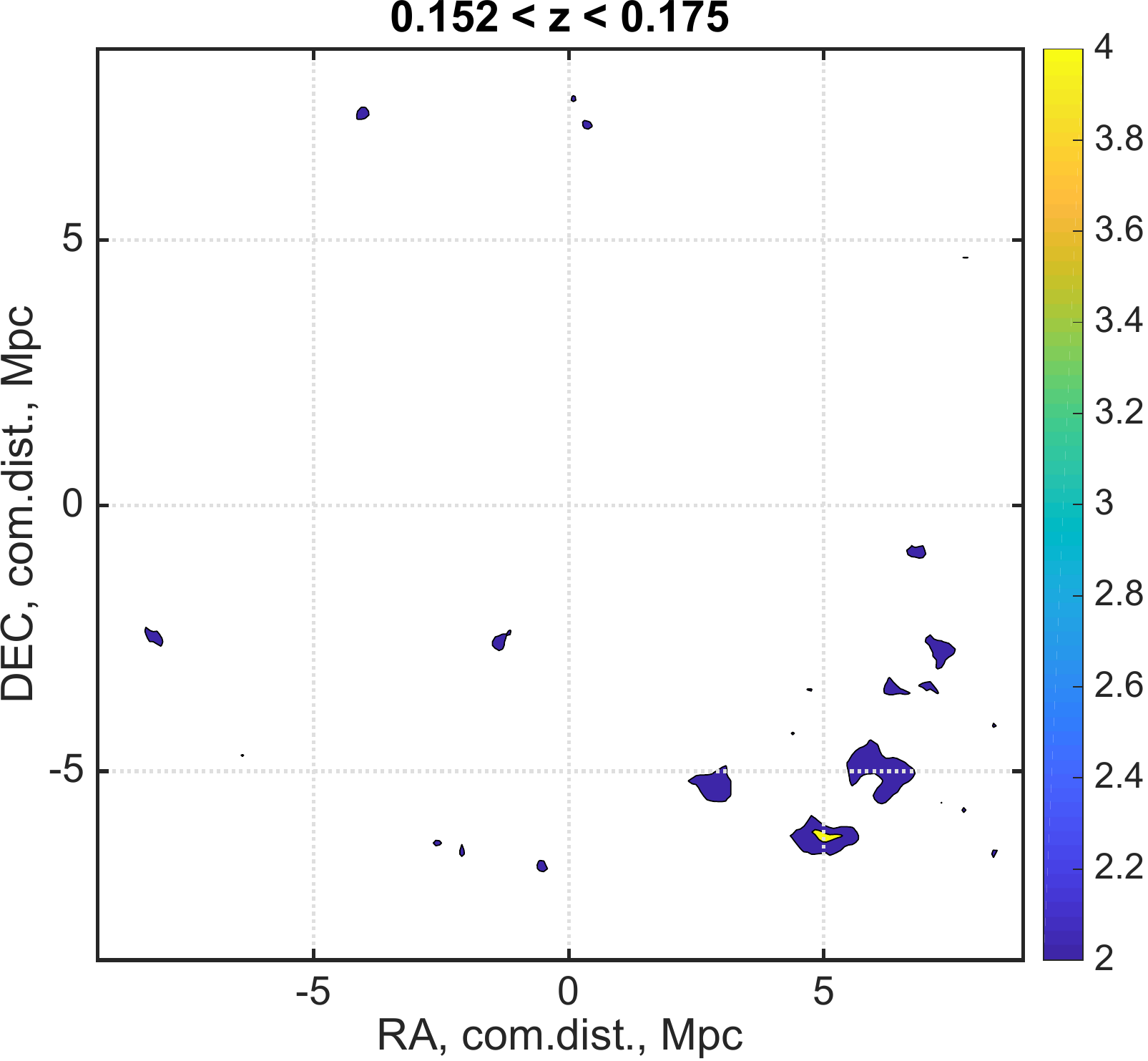}} \\  
 \end{minipage}
  \hfill
\begin{minipage}[h]{0.45\linewidth} 
\center{\includegraphics[scale=0.45]{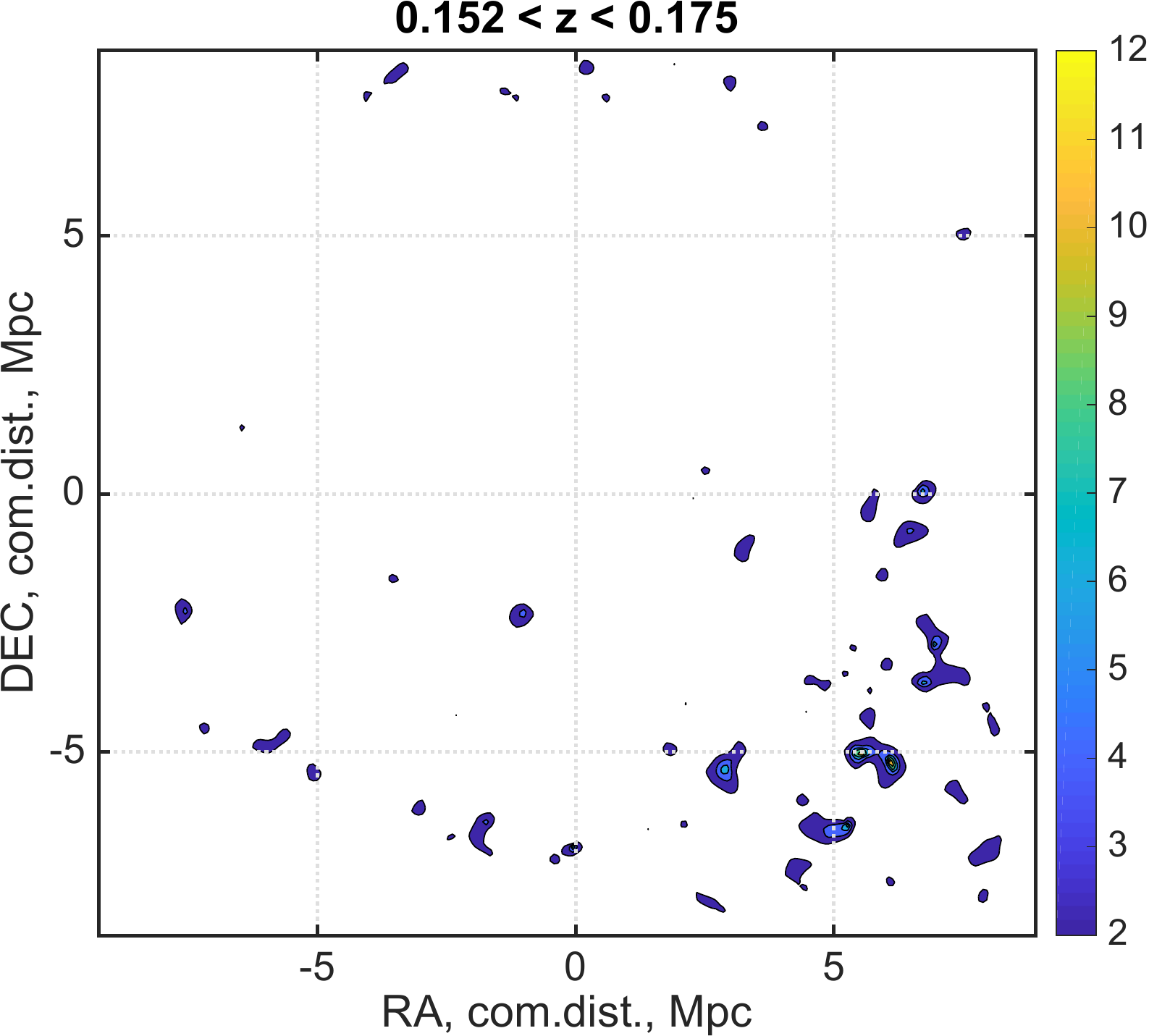}} \\  
 \end{minipage}
 \caption{Density contrast maps reconstructed for every fifth layer using the Voronoi tessellation (right) and the adaptive aperture method with smoothing the density of the environment (left). The color bar shows how many times the density of the environment is exceeded in comparison with the average density in the layer. The axes are the coordinates of right ascension (RA) and declination (DEC), expressed in Mpc (comoving distance). A complete set of 57 redshift slices is available on \href{https://github.com/ale-gro/density_maps}{https://github.com/ale-gro/densitymaps}.}
  \label{fig_10}
\end{figure*}

 \begin{figure*}[]
\begin{minipage}[h]{0.45\linewidth} 
\center{\includegraphics[scale=0.45]{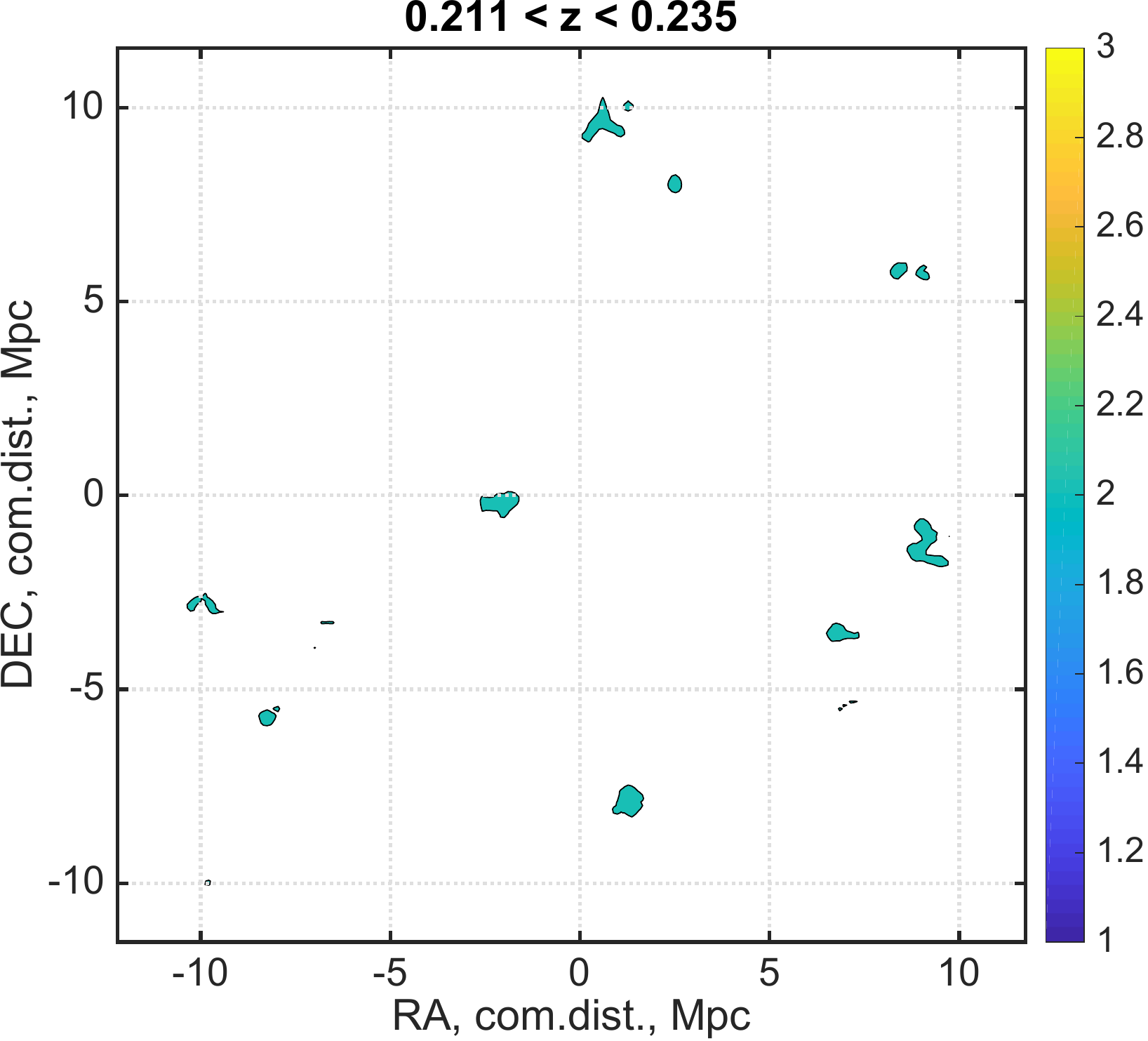}}  \\  
 \end{minipage}
 \hfill
\begin{minipage}[h]{0.45\linewidth}  
\center{\includegraphics[scale=0.45]{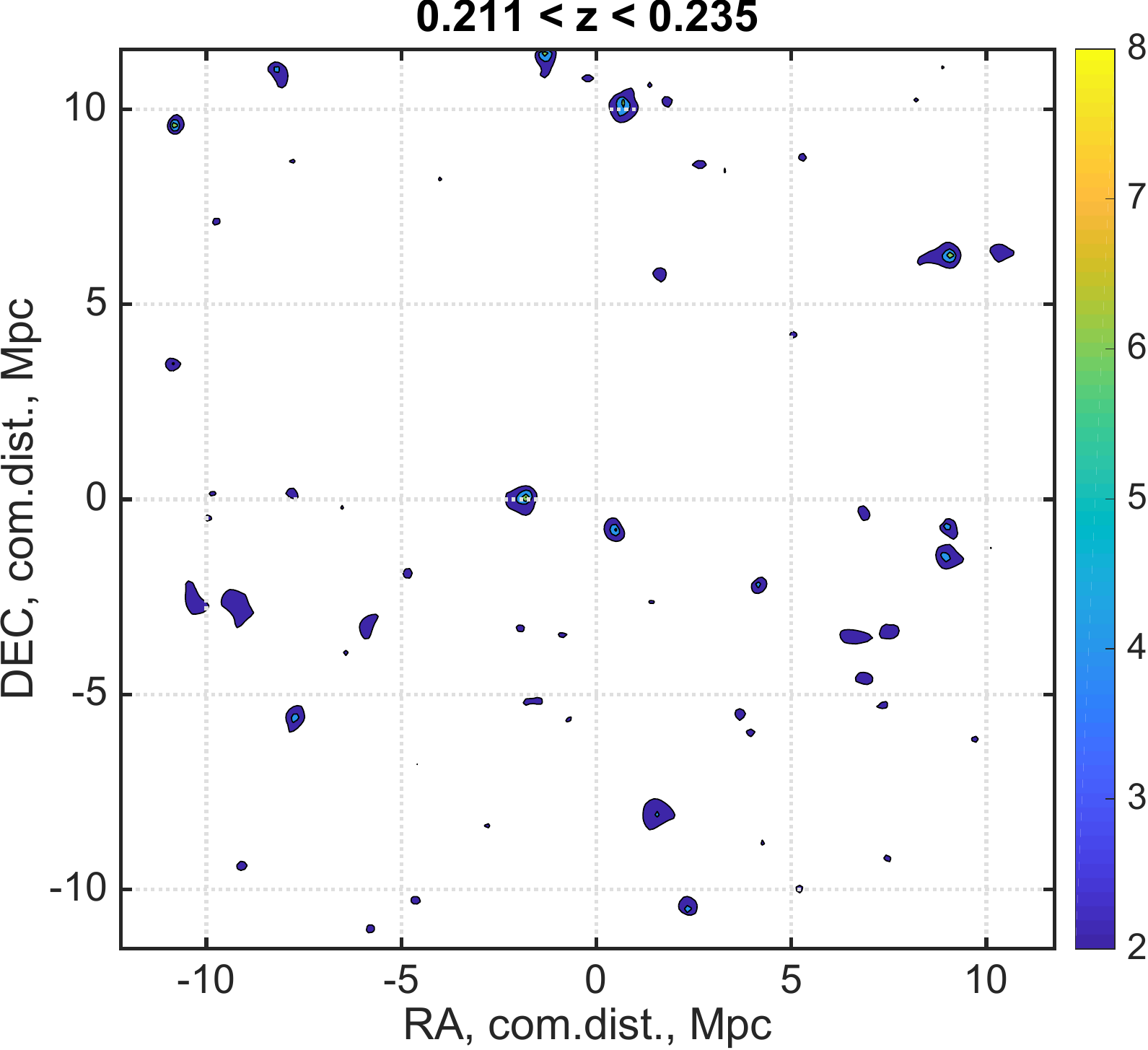}}  \\  
 \end{minipage}
 \vfill
 \begin{minipage}[h]{0.45\linewidth}  
\center{ \includegraphics[scale=0.45]{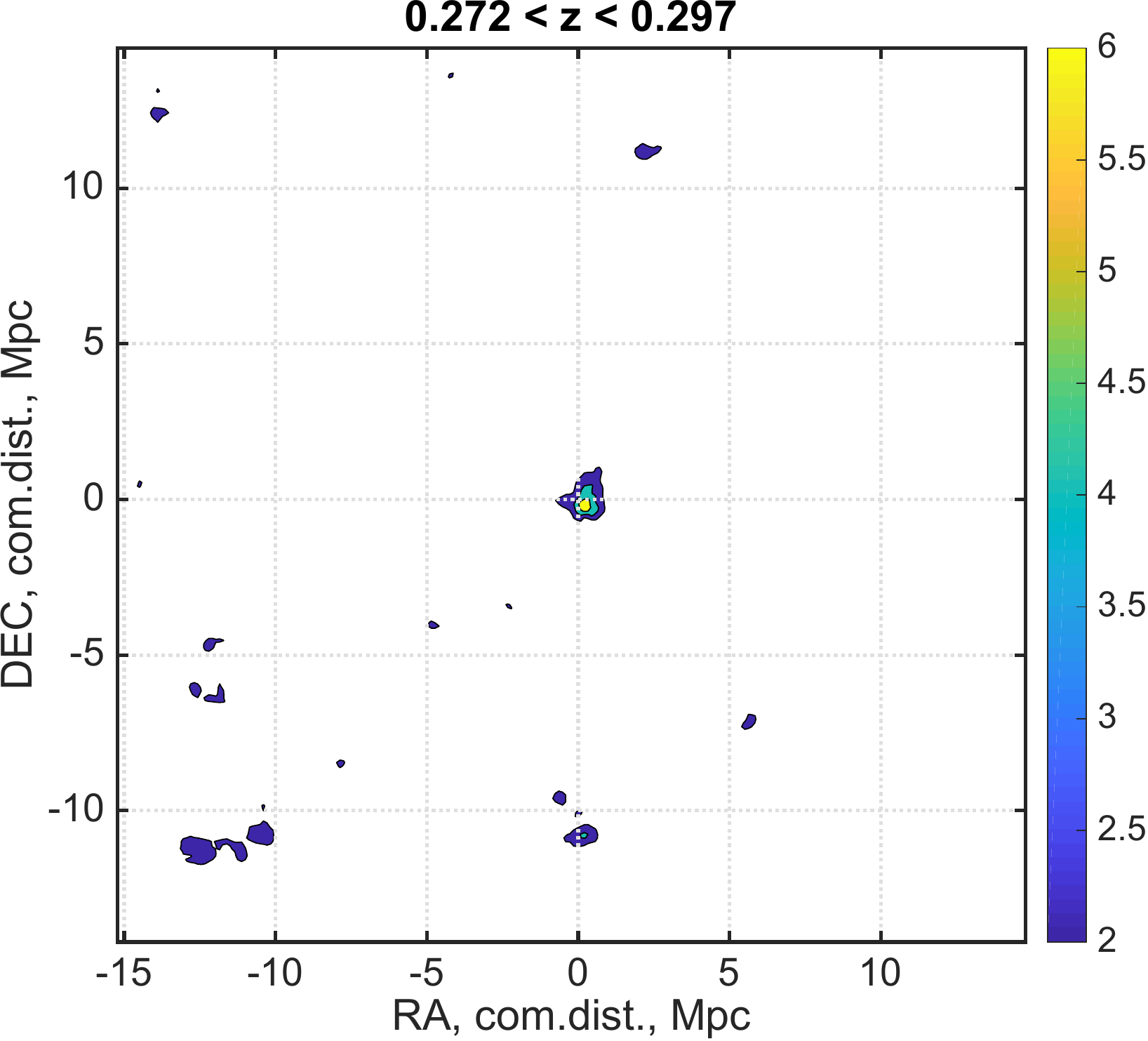}} \\  
 \end{minipage}
  \hfill
\begin{minipage}[h]{0.45\linewidth} 
\center{\includegraphics[scale=0.45]{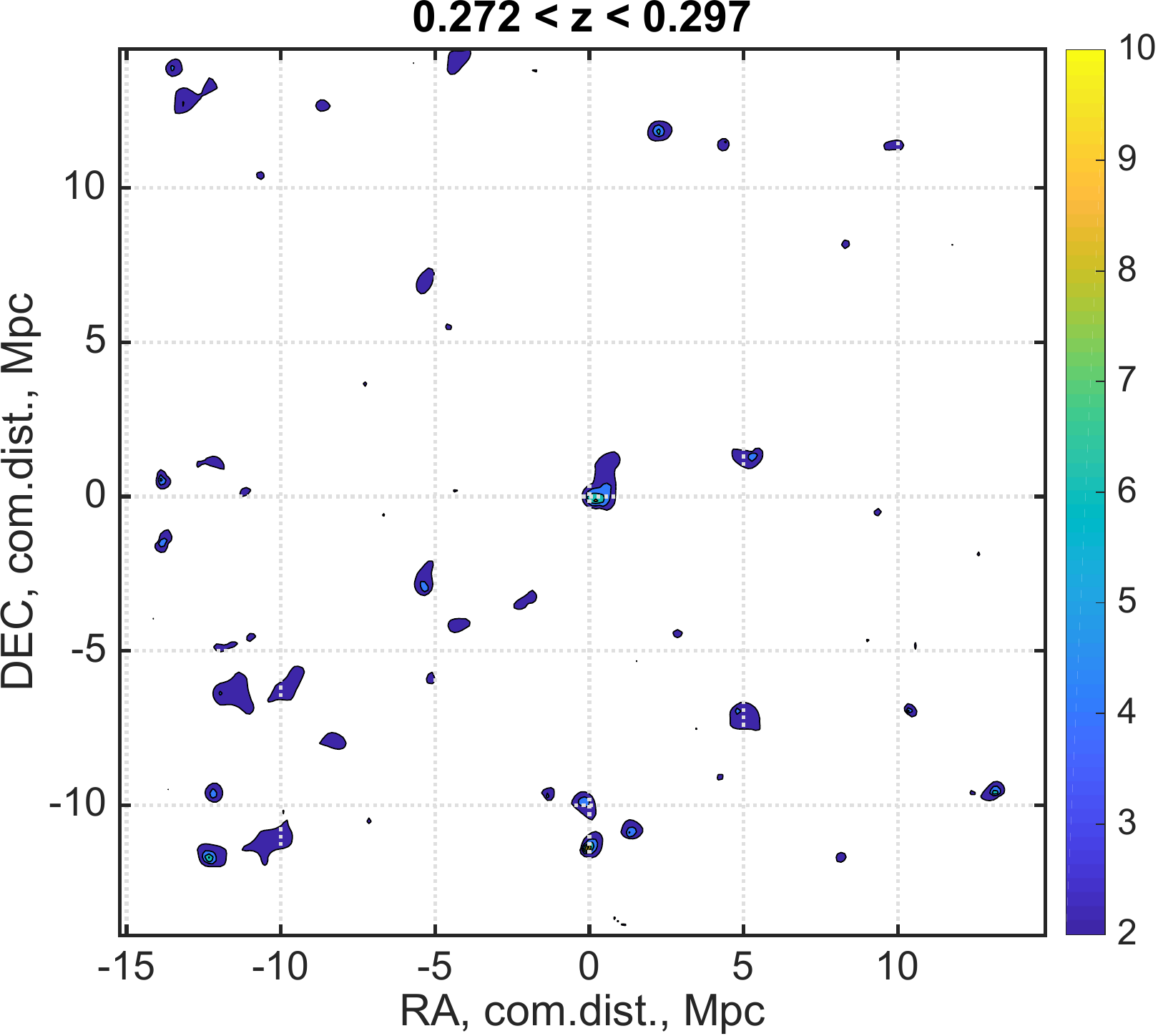}} \\  
 \end{minipage}
  \vfill
 \begin{minipage}[h]{0.45\linewidth}  
\center{ \includegraphics[scale=0.45]{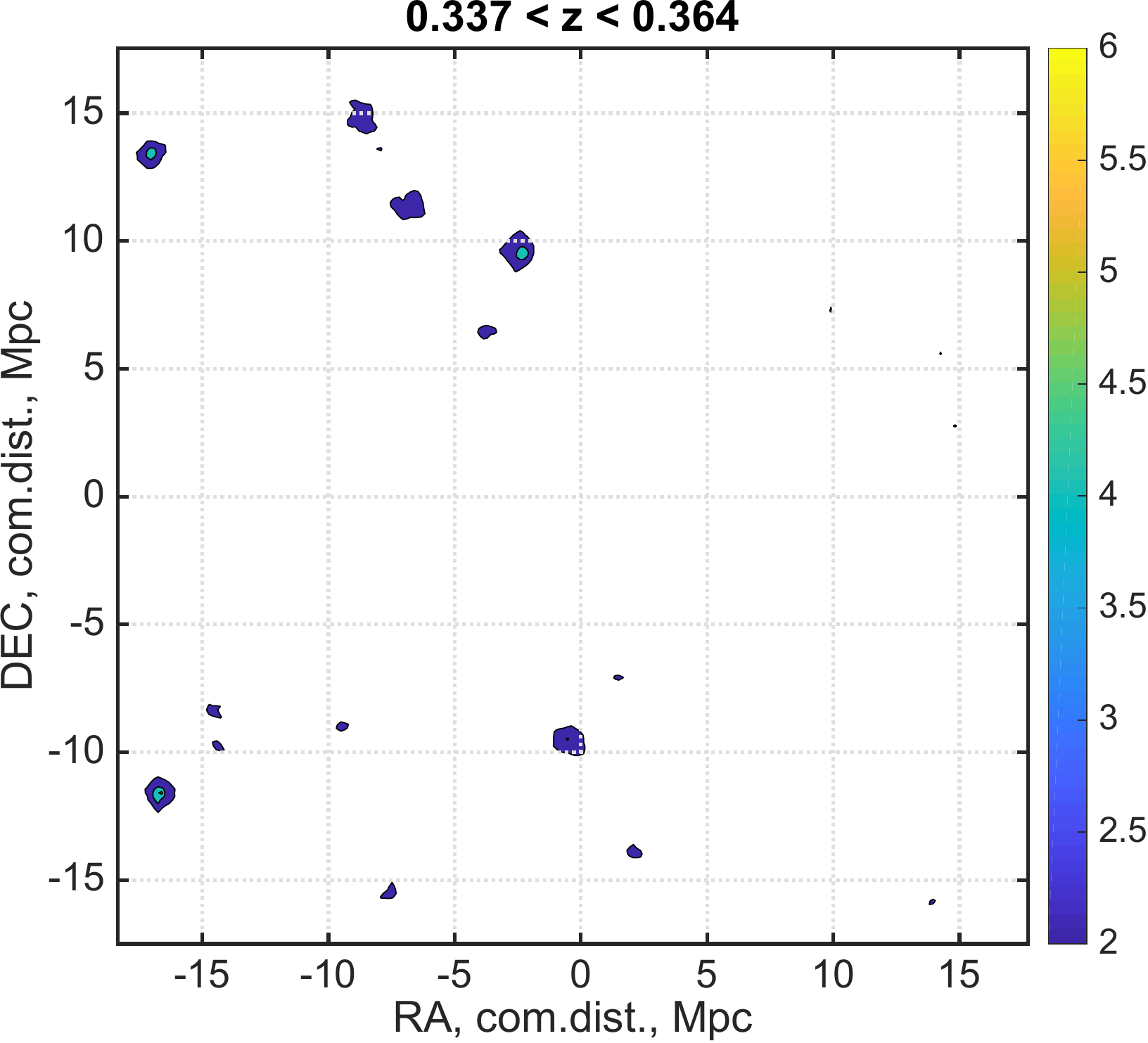}} \\  
 \end{minipage}
  \hfill
\begin{minipage}[h]{0.45\linewidth} 
\center{\includegraphics[scale=0.45]{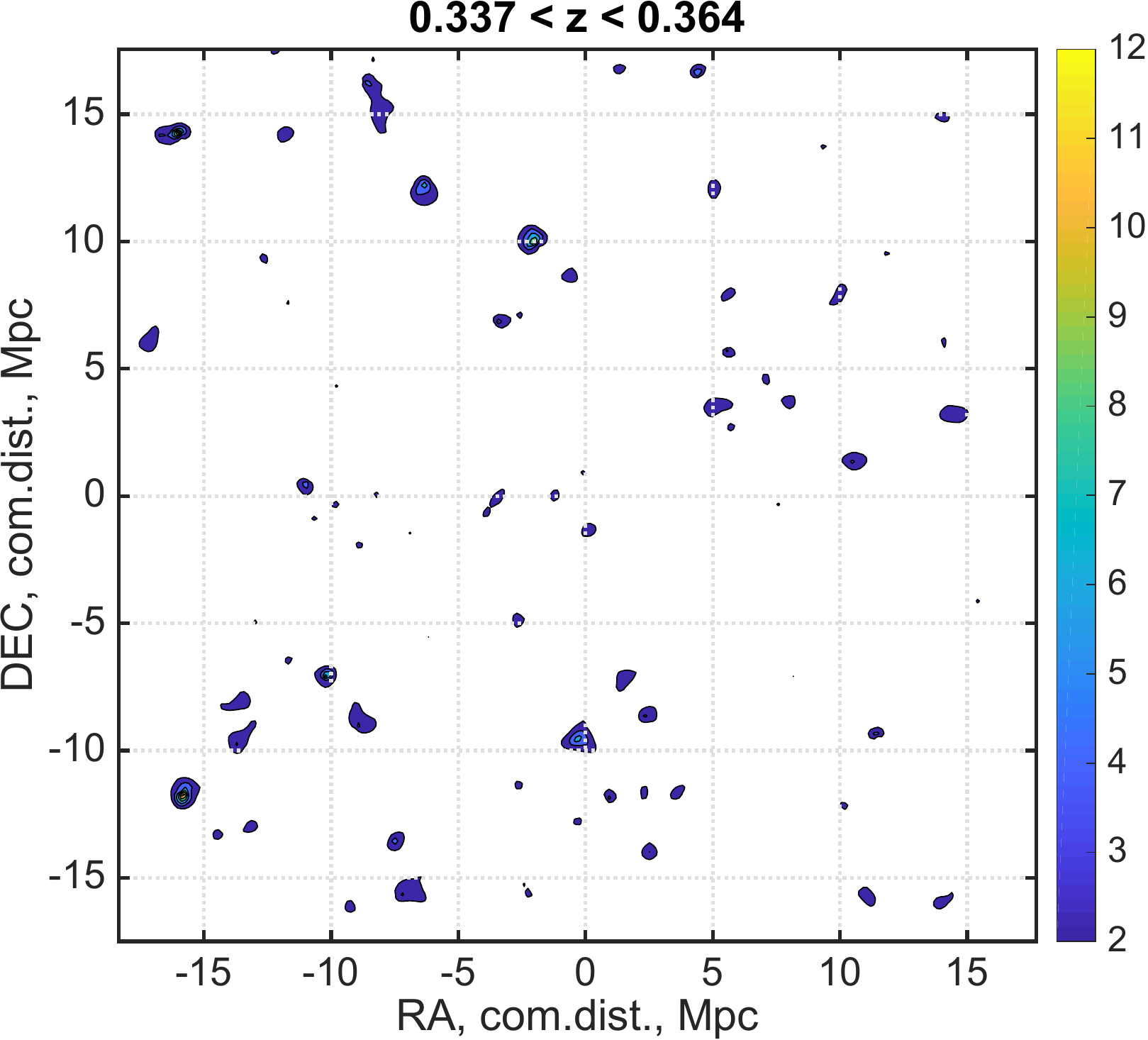}} \\  
 \end{minipage}
 
 \figurename{ \textbf{10.} Continued.}
  \label{}
\end{figure*}

 \begin{figure*}[]
\begin{minipage}[h]{0.45\linewidth} 
\center{\includegraphics[scale=0.45]{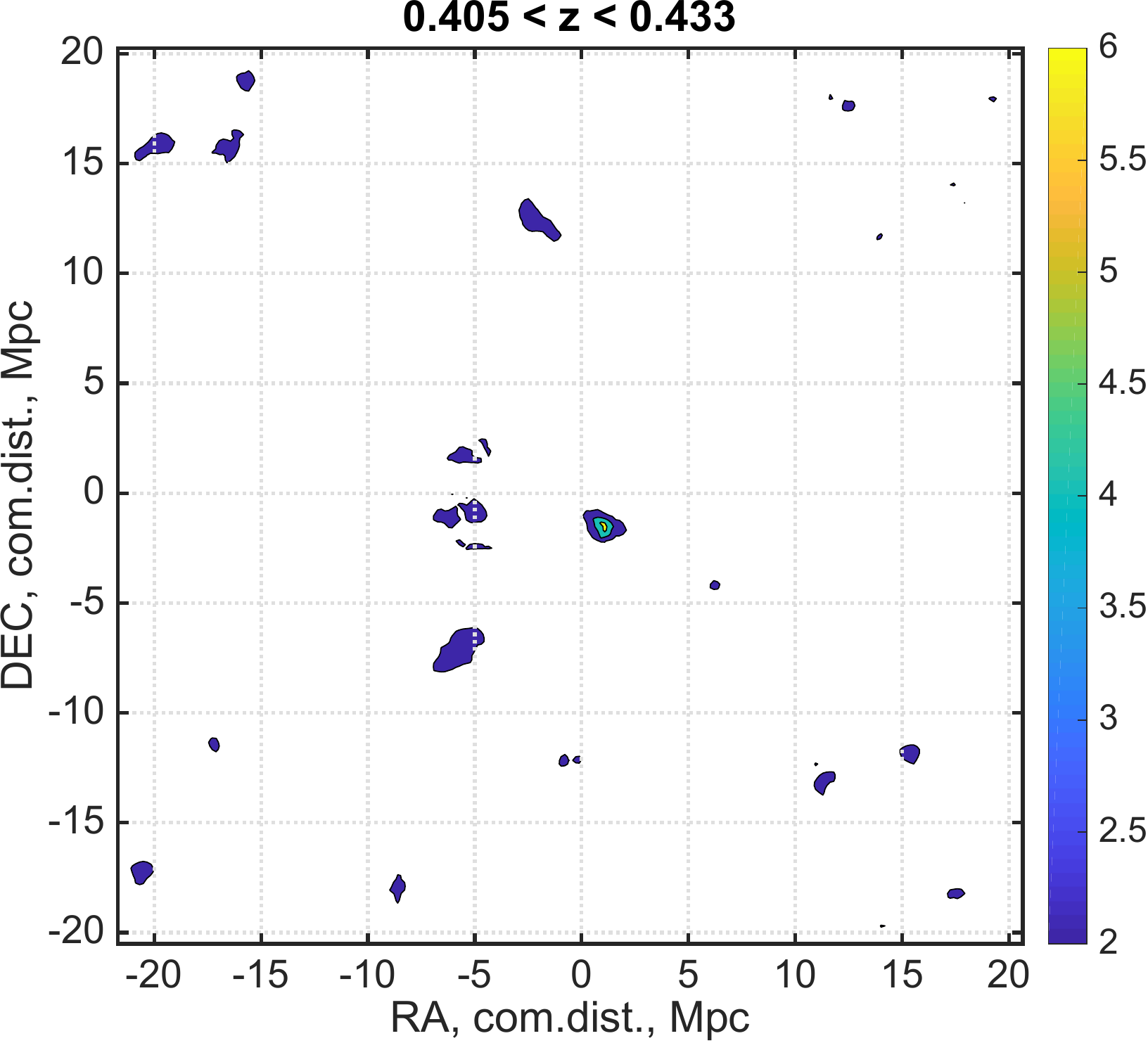}}  \\  
 \end{minipage}
 \hfill
\begin{minipage}[h]{0.45\linewidth}  
\center{\includegraphics[scale=0.45]{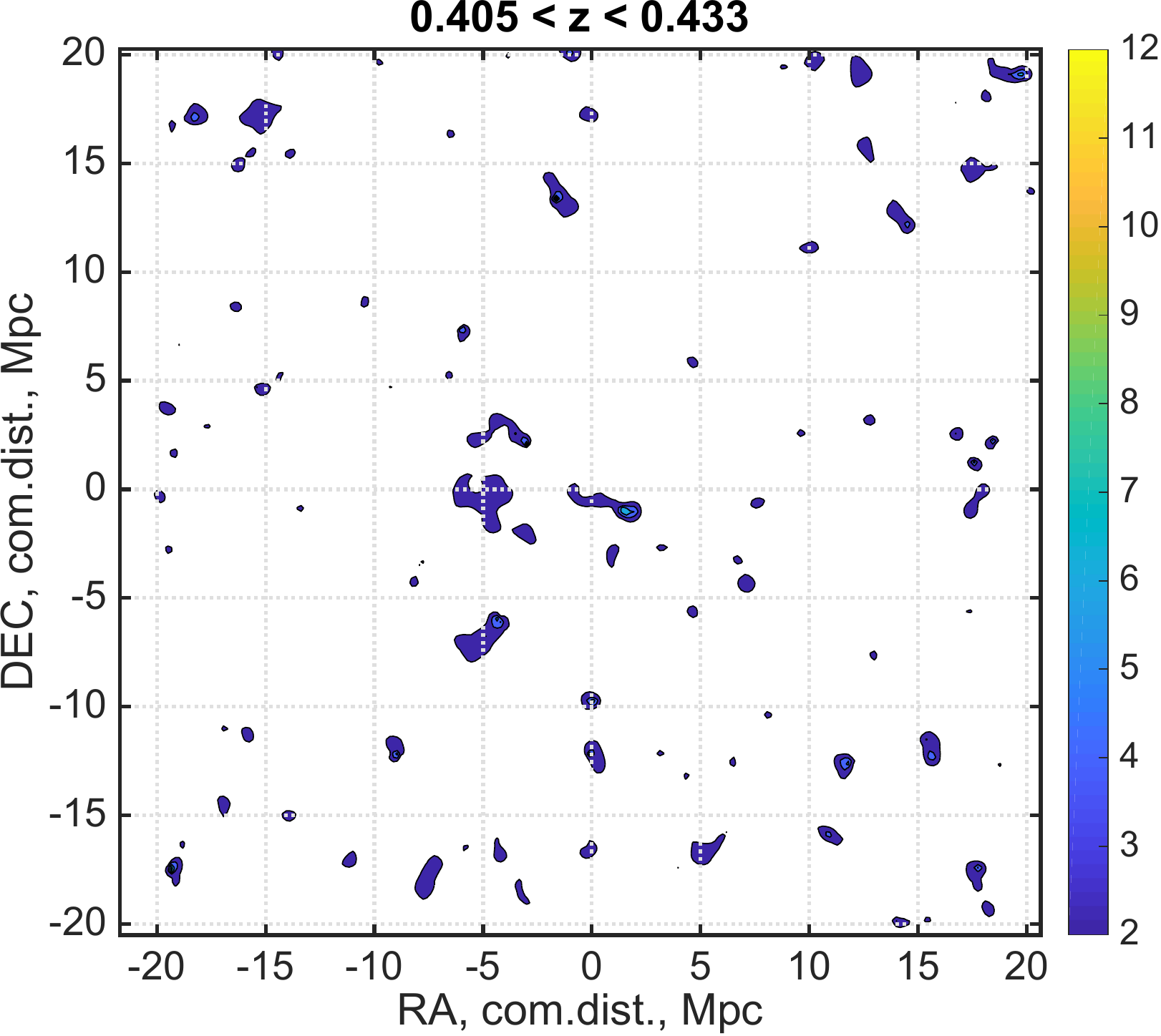}}  \\  
 \end{minipage}
 \vfill
 \begin{minipage}[h]{0.45\linewidth}  
\center{ \includegraphics[scale=0.45]{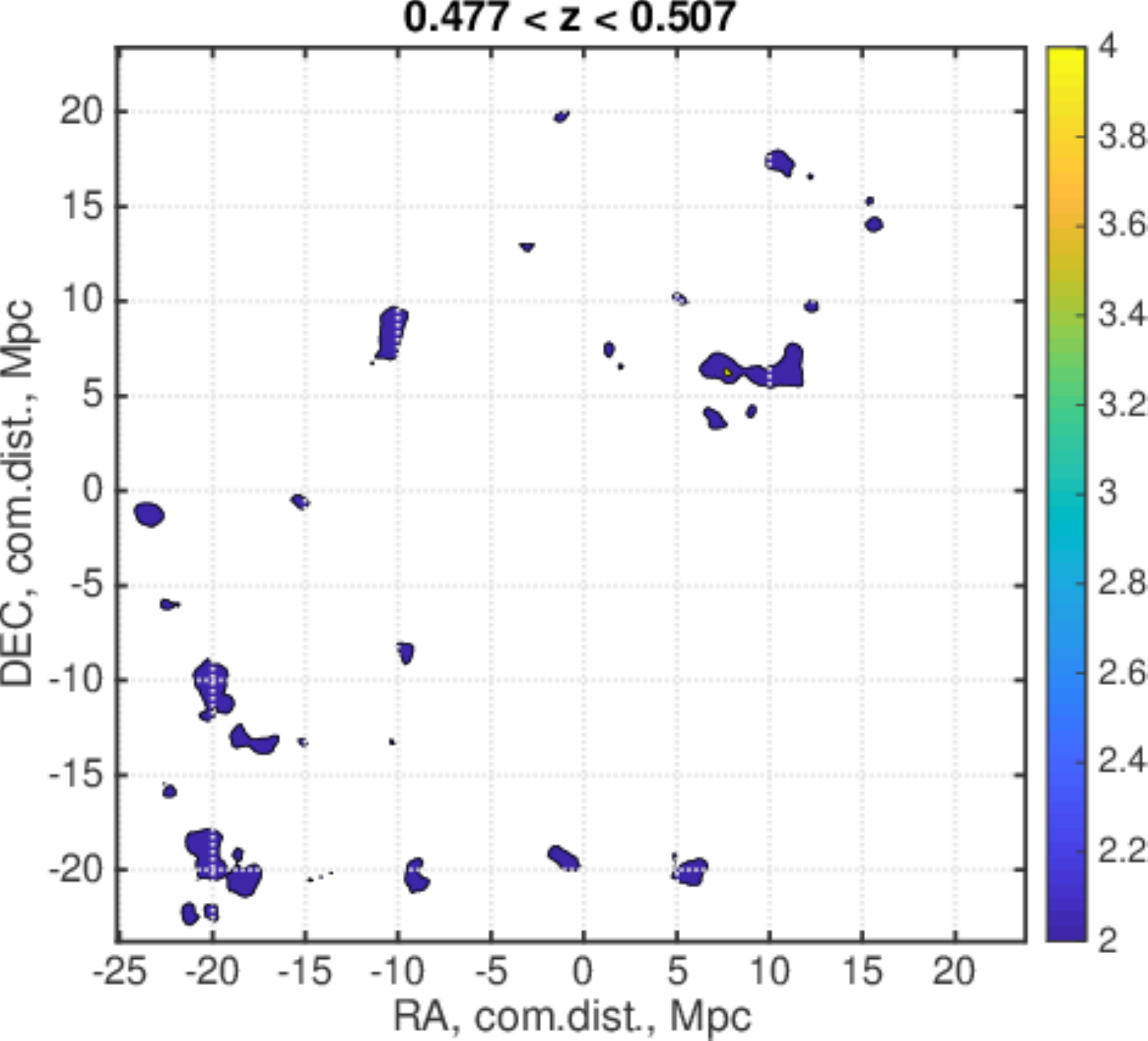}} \\  
 \end{minipage}
  \hfill
\begin{minipage}[h]{0.45\linewidth} 
\center{\includegraphics[scale=0.45]{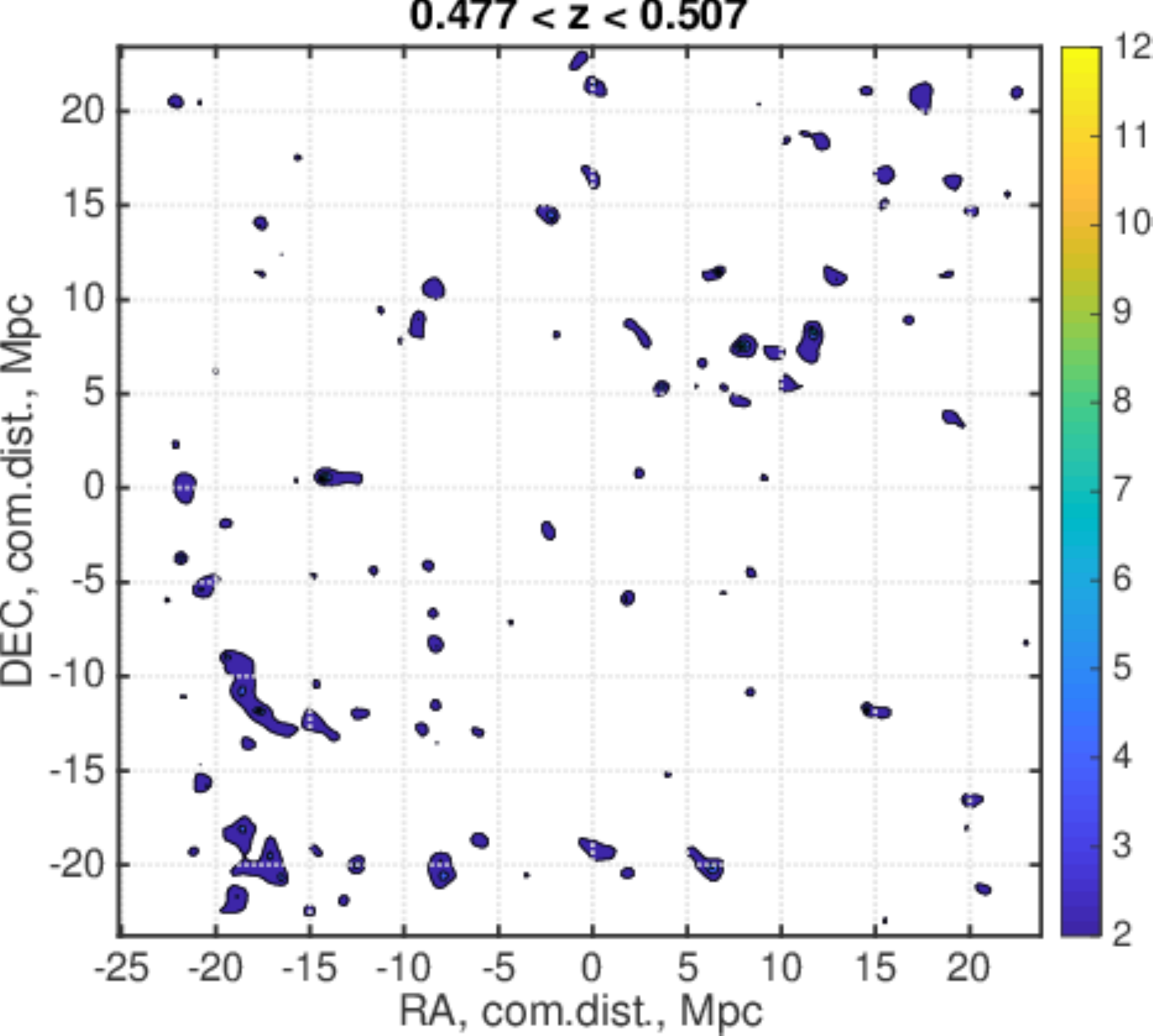}} \\  
 \end{minipage}
  \vfill
 \begin{minipage}[h]{0.45\linewidth}  
\center{ \includegraphics[scale=0.45]{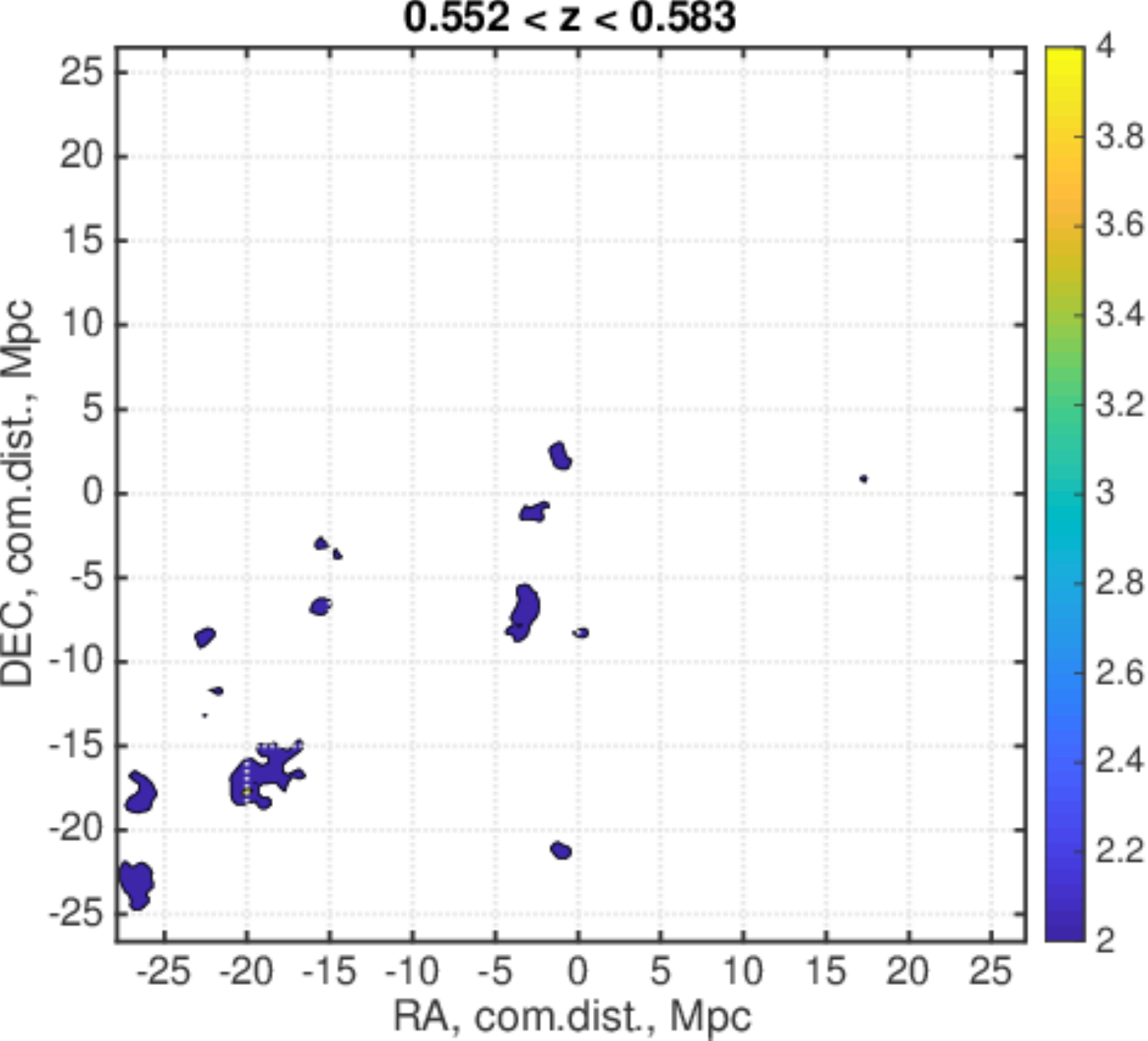}} \\  
 \end{minipage}
  \hfill
\begin{minipage}[h]{0.45\linewidth} 
\center{\includegraphics[scale=0.45]{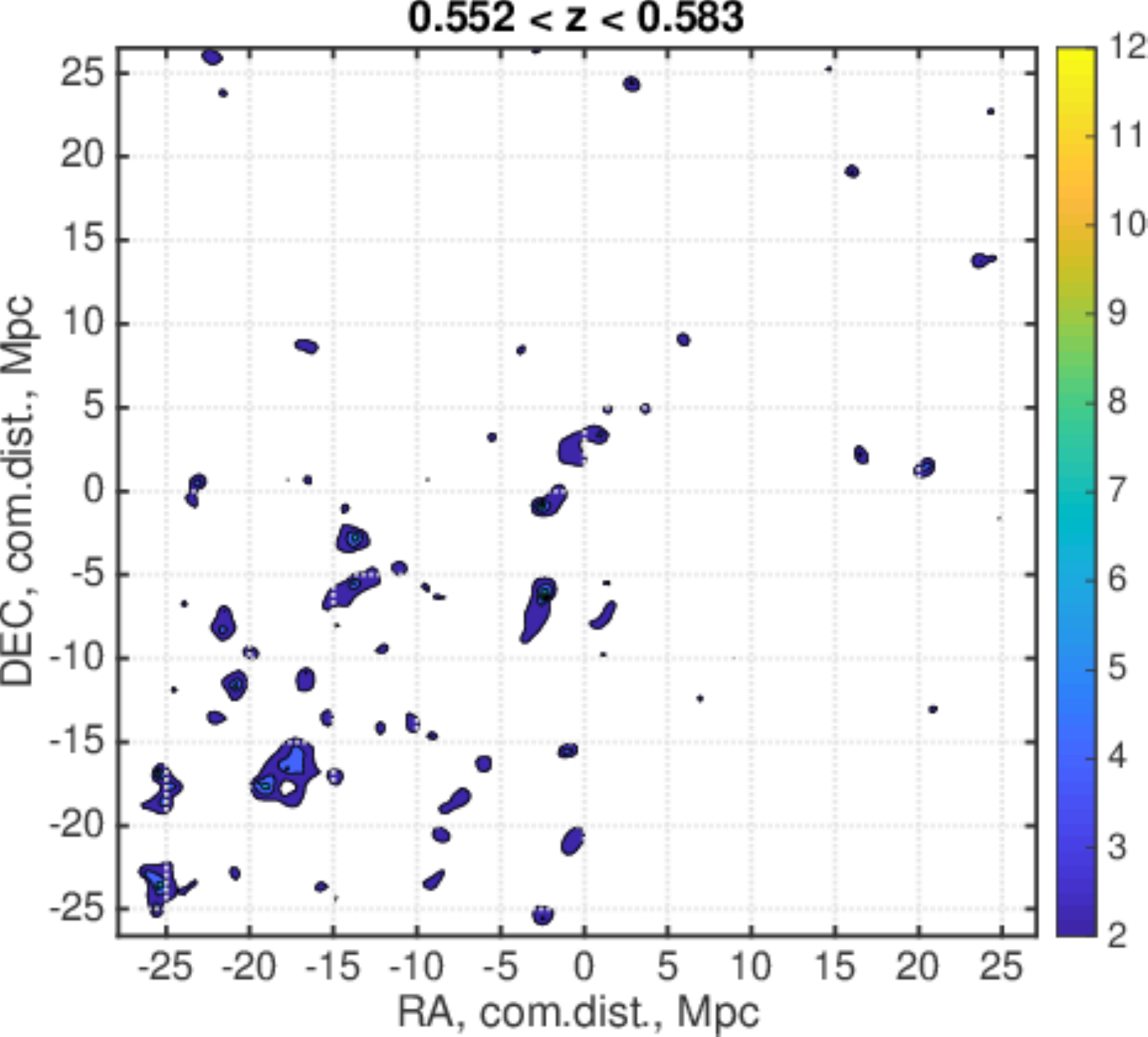}} \\  
 \end{minipage}
 
 \figurename{ \textbf{10.} Continued.}
  \label{}
\end{figure*}

 \begin{figure*}[]
\begin{minipage}[h]{0.45\linewidth} 
\center{\includegraphics[scale=0.45]{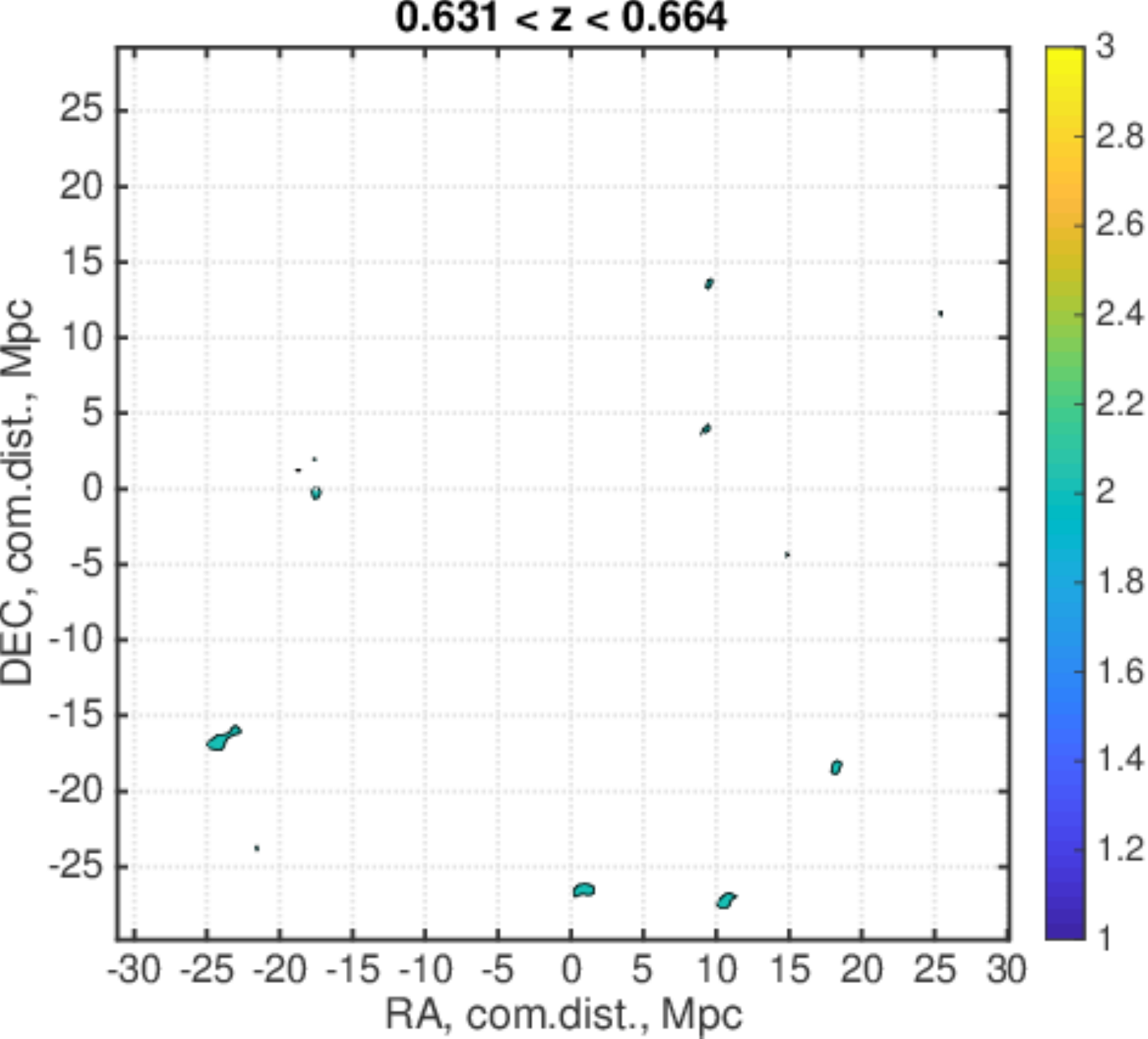}}  \\  
 \end{minipage}
 \hfill
\begin{minipage}[h]{0.45\linewidth}  
\center{\includegraphics[scale=0.45]{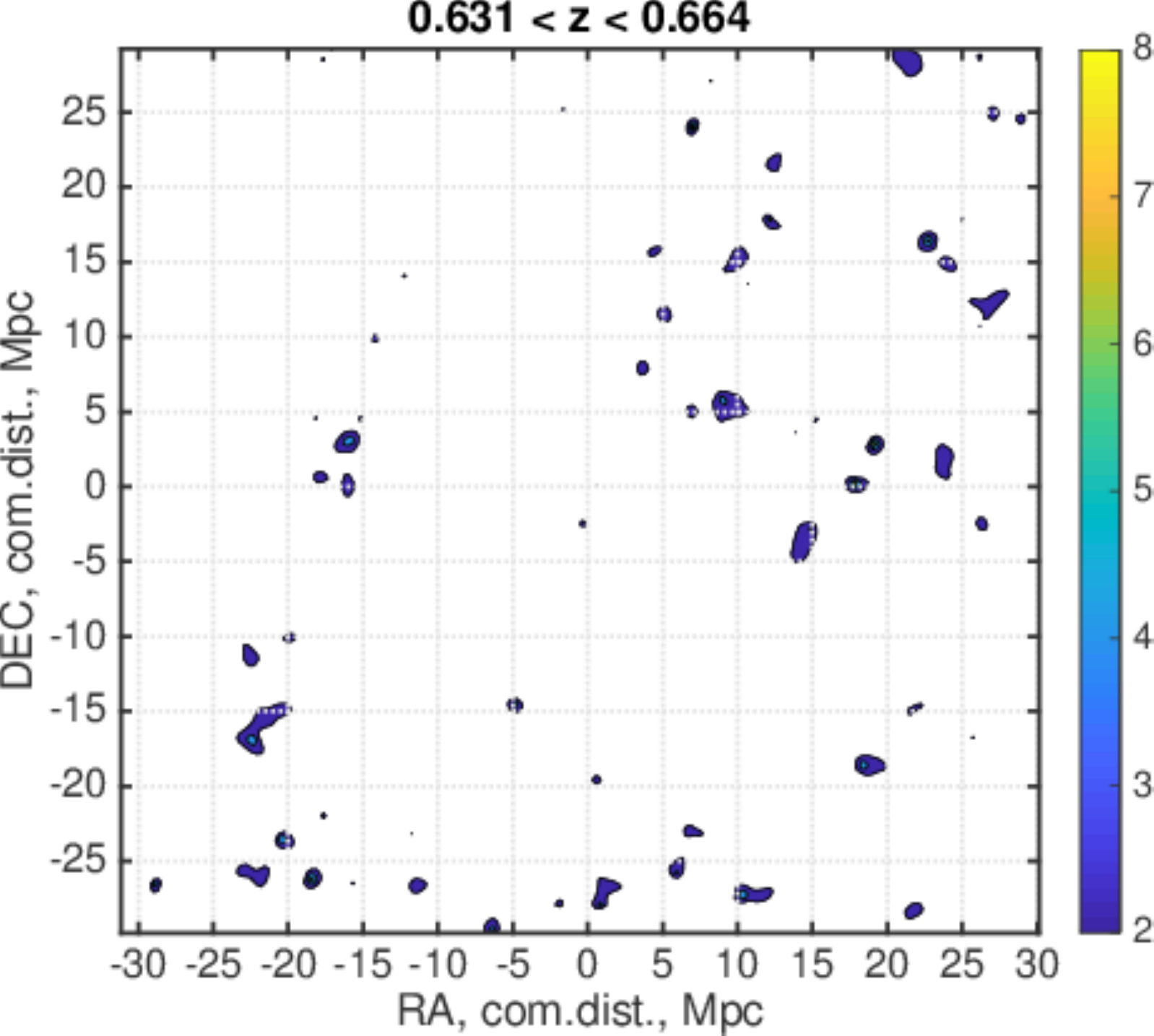}}  \\  
 \end{minipage}
 \vfill
 \begin{minipage}[h]{0.45\linewidth}  
\center{ \includegraphics[scale=0.45]{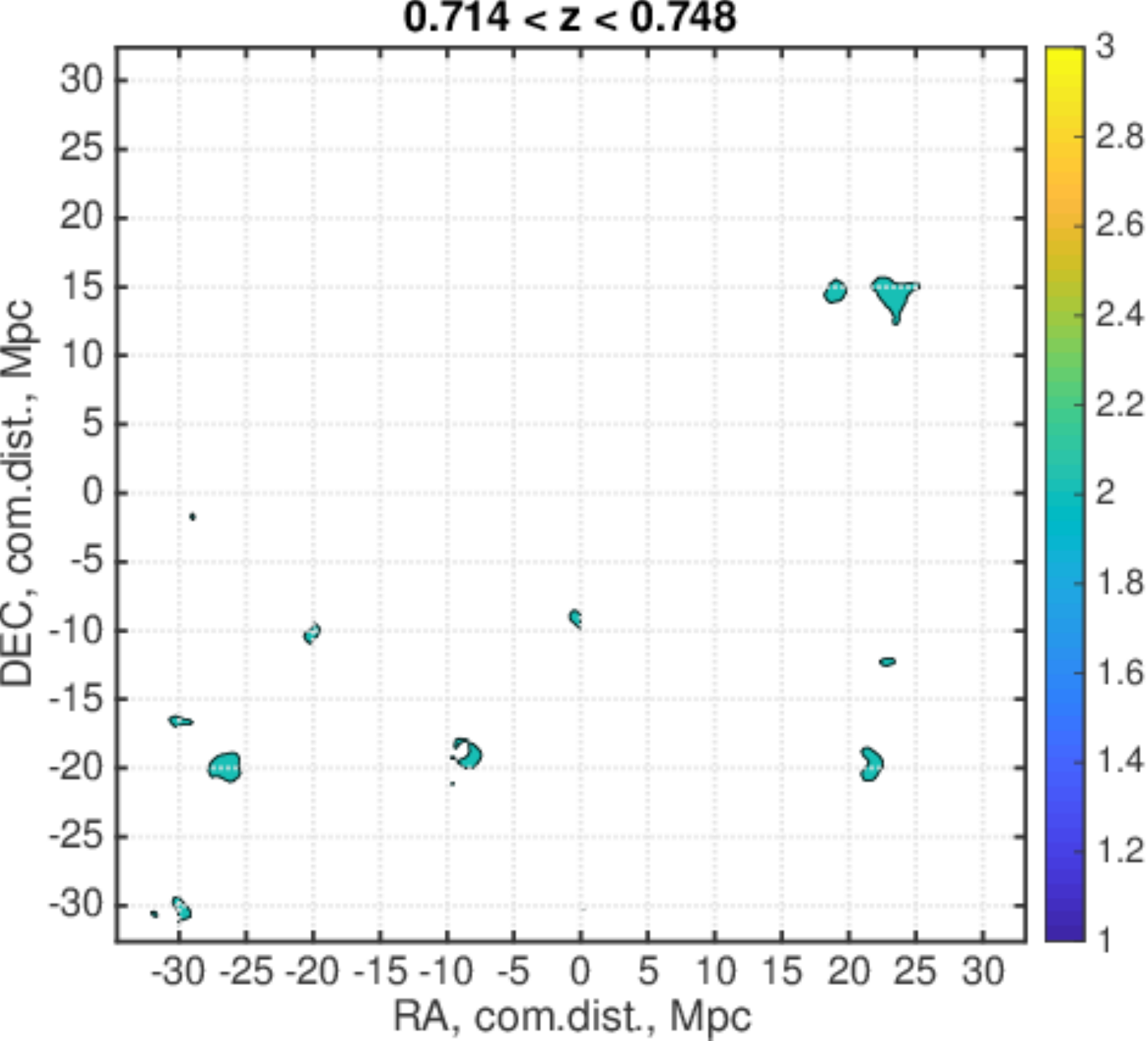}} \\  
 \end{minipage}
  \hfill
\begin{minipage}[h]{0.45\linewidth} 
\center{\includegraphics[scale=0.45]{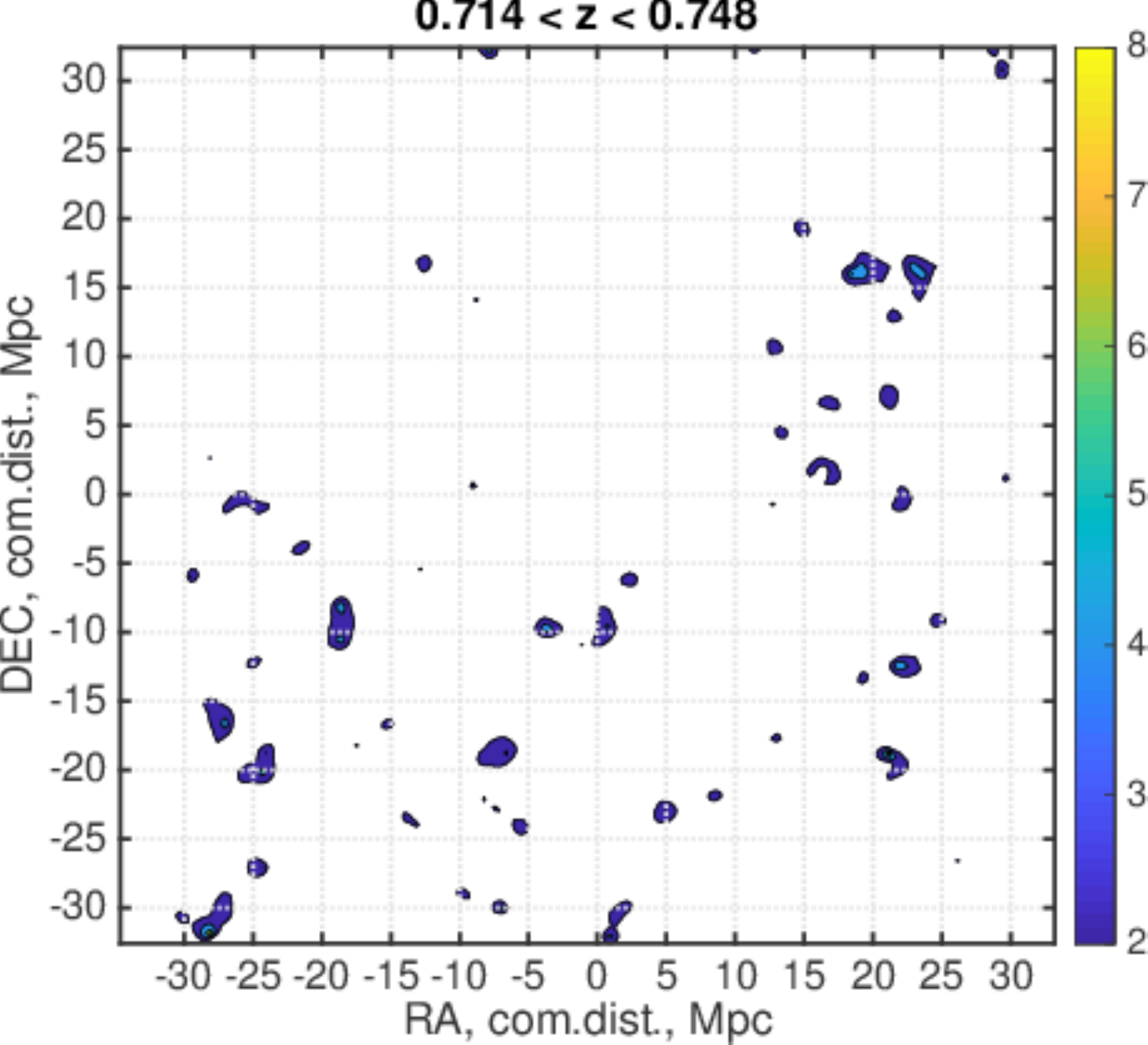}} \\  
 \end{minipage}
 
 \figurename{ \textbf{ 10.} Ending.}
  \label{}
\end{figure*}

Fig. 5 shows the Voronoi tessellations for each galaxy (red for galaxies with early SED types E - Sa, green for late SED types Sab - Sd, and blue for irregular galaxies IRR / starburst galaxies SB) for a slice of $ 0.477 \leq z \leq 0.507 $. In Fig. 6 (right) shows the density contrast maps obtained from the analysis of Voronoi tessellations for the same slice and in Fig. 10 (right) of the density contrast map for every fifth slice at redshift. Similarly, in Fig. 6 (left) and in Fig. 10 (left) - contrast maps reconstructed using the adaptive aperture and smoothing method. These figures illustrate well the spatial overdensities in the distribution of galaxies. The density contrast maps obtained show a wide range of large-scale overdensities from 0.5 to 10 Mpc (concomitant size) over the full range of redshifts. Overdensities demonstrate various spatial forms from almost regular round to elongated filaments.

Large galaxy samples available from photometric redshift catalogs allow structures to be mapped even at relatively low densities. We found more than 250 regions with high-density contrast on scales from 0.5 Mpc to 10 Mpc (comoving size). A catalog of clusters and clusters of galaxies will be published in \citep {Grokhovskaya2020}.

\begin{figure*}[]
\begin{minipage}[h]{0.4\linewidth} 
\center{\includegraphics[scale=0.5]{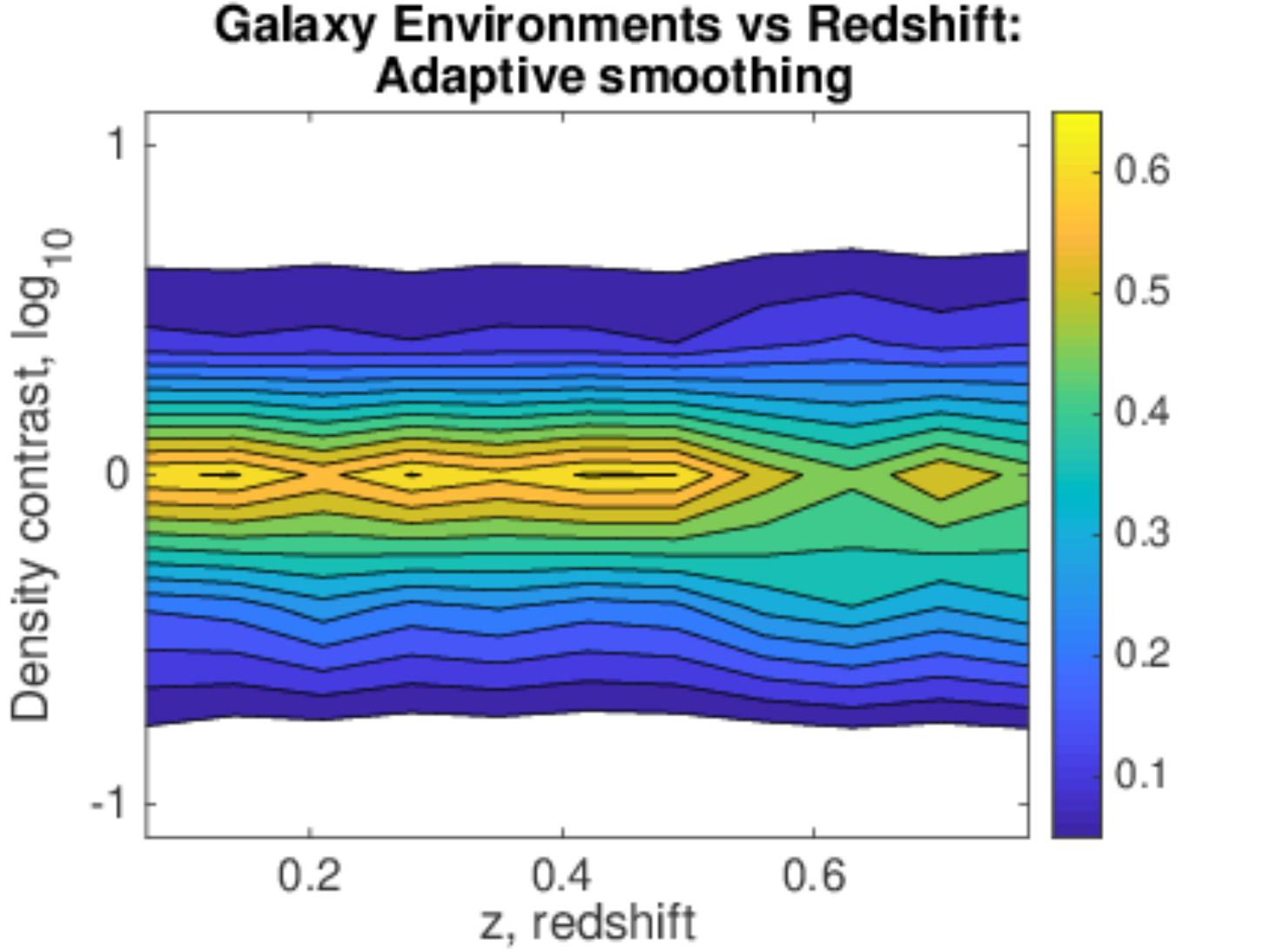}}\\ 
 \end{minipage}
 \hfill
\begin{minipage}[h]{0.4\linewidth}  
\center{ \includegraphics[scale=0.5]{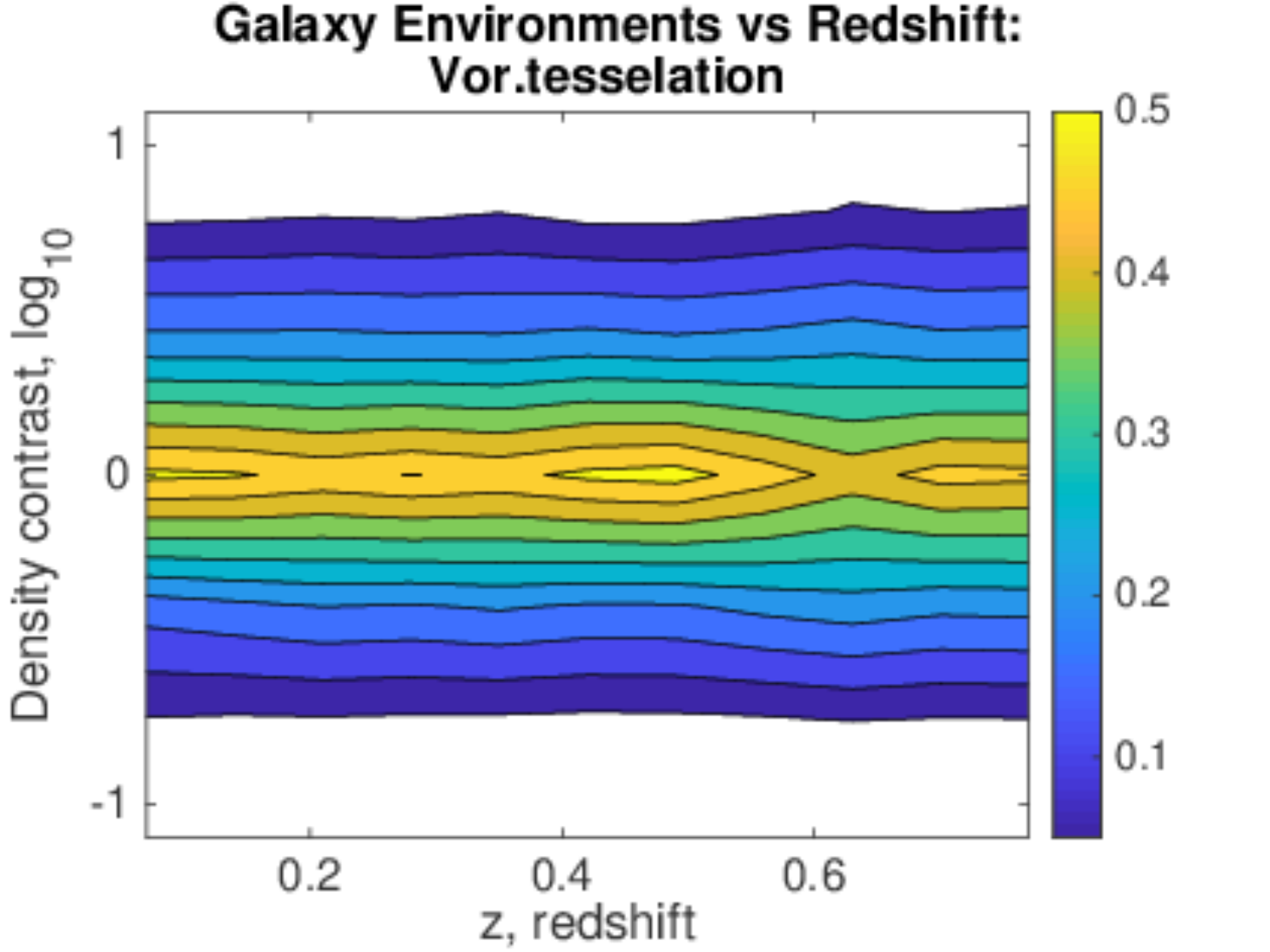}}\\
 \end{minipage}
 \caption{Density contrast obtained by the adaptive aperture and smoothing method for 28,398 galaxies (left) and the Voronoi tessellation for 27,446 galaxies (right) from the photometric catalog in the redshift range from 0 to 0.8. The color shows the contours for the relative number of galaxies of each density contrast value greater than 0.05 and each interval by redshift. Throughout the entire range of redshifts, the bulk of the galaxies are in areas with an average density of the environment in the layer.}
\label{fig_11}
\end{figure*}

Fig. 11 shows the full range of density contrast values for the environment of galaxies from the studied sample, depending on the redshift. The outlines show the relative number of galaxies as a function of contrast in the density of the environment and redshift. In general, between the values obtained by two independent methods for determining the density, a good agreement is seen between the results: both in the relative number of galaxies at different environmental densities, and in a change in the structure of the density contrast with redshift. However, the set of density contrast values for each of the sample galaxies obtained by the adaptive aperture and smoothing method contains fewer overdense and underdense regions. This is because this algorithm extracts only statistically significant structures.

\begin{figure*}[]
\begin{minipage}[h]{0.4\linewidth} 
\center{\includegraphics[scale=0.5]{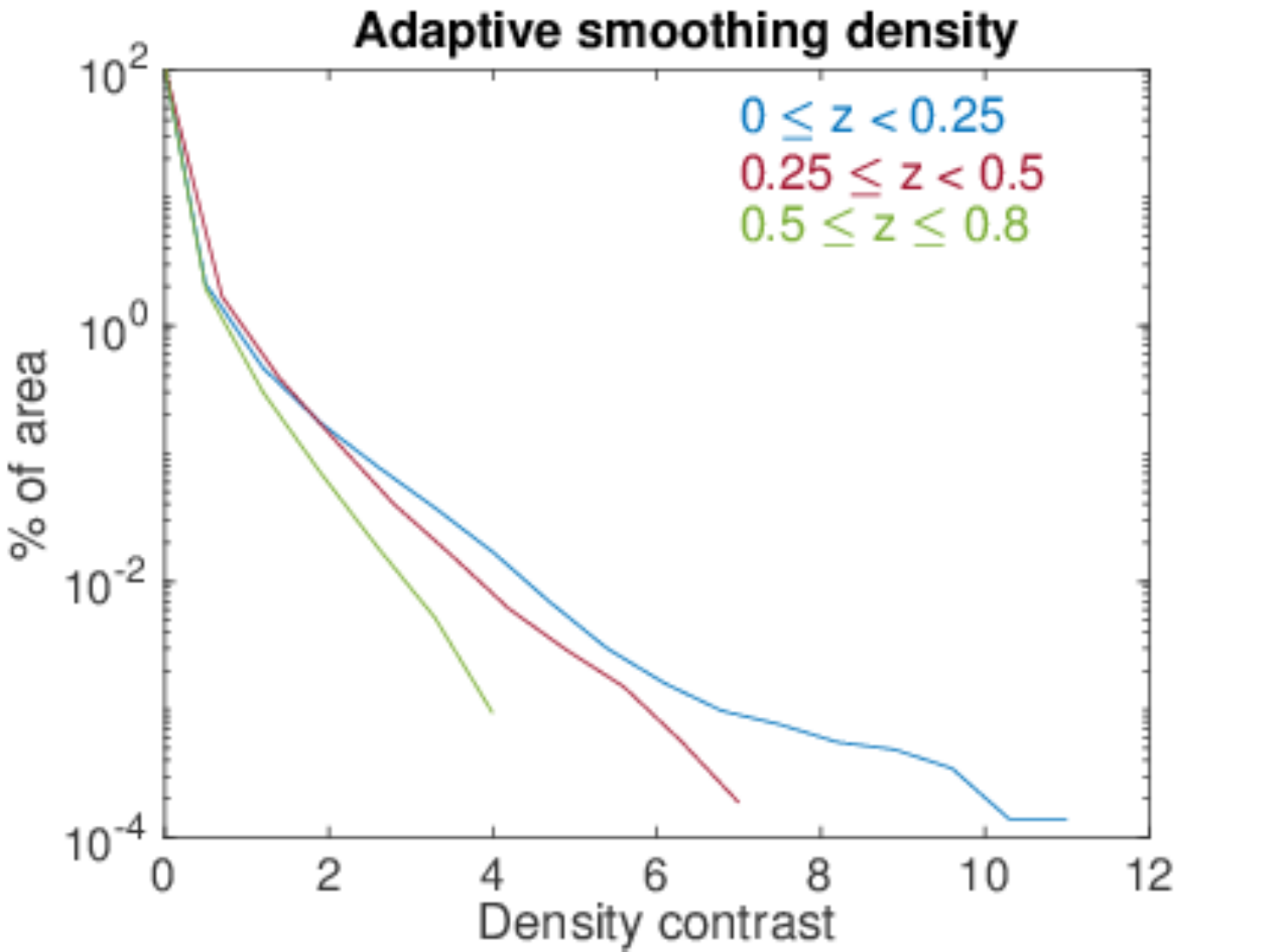}}\\ 
 \end{minipage}
 \hfill
\begin{minipage}[h]{0.4\linewidth}  
\center{ \includegraphics[scale=0.5]{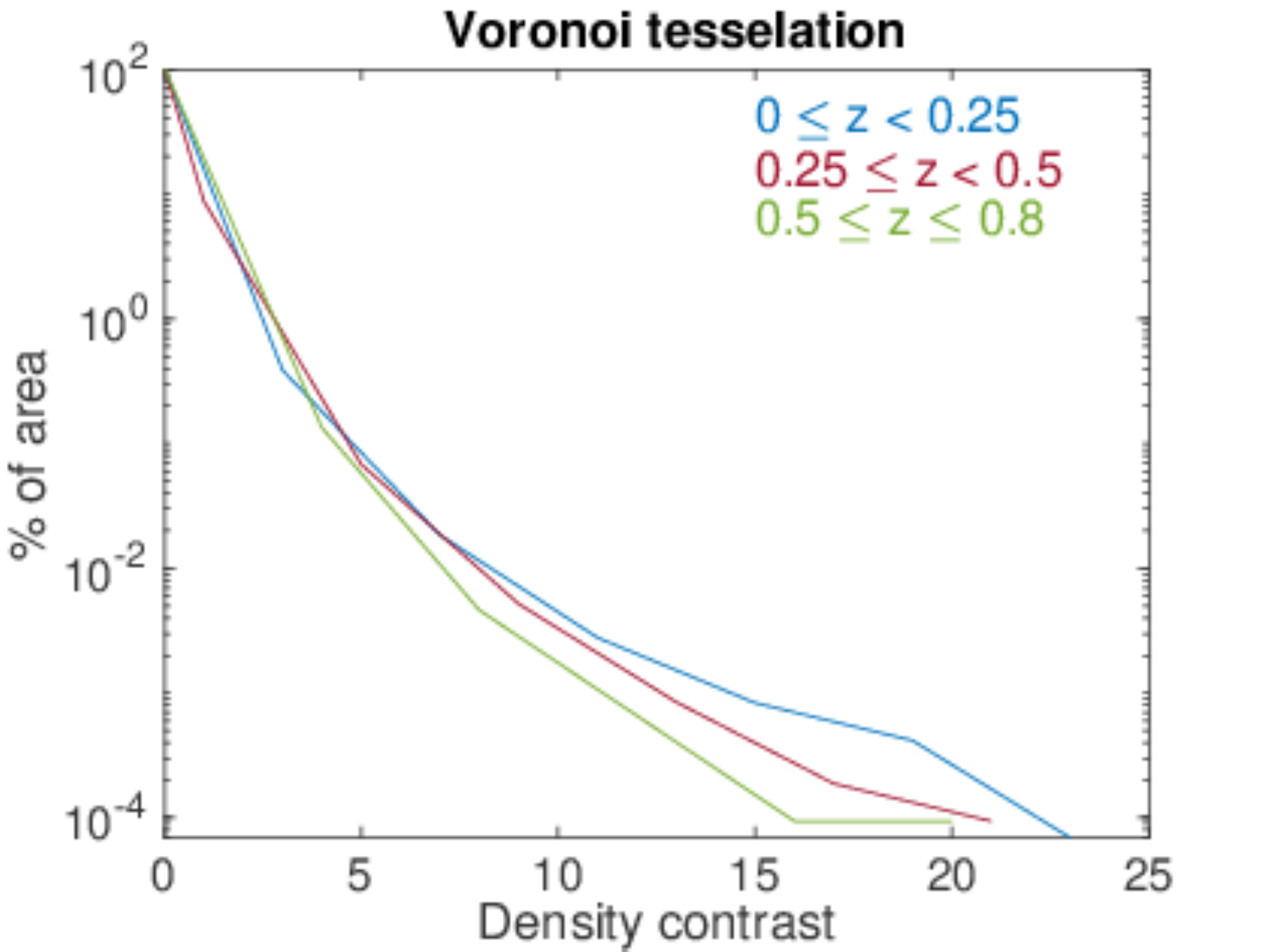}}\\
 \end{minipage}
 \caption{The percentage of the area of the field is shown depending on the value of the contrast of the density of the environment, for three ranges of redshift for two methods for reconstructing density contrast maps: the method with adaptive aperture and smoothing of the density of the environment (left) and the method of Voronoi tessellation (right). A different range along the axis of the contrast of the density of the environment is caused by differences in the operation of the algorithms}
\label{fig_12}
\end{figure*}

In Fig. 12 shows the percentage of the field area as a function of density contrast for the three redshift ranges ($ 0 \leq z <0.25 $, $ 0.25 \leq z <0.5 $, $ 0.5 \leq z \leq 0.8 $) for both methods of restoring contrast maps. The figure clearly shows an increase in the range of density contrast for high redshifts ($ 0.5 \leq z \leq 0.8 $) compared to those observed at low redshifts ($ 0 \leq z <0.25 $) for both methods. The difference in the magnitude of the ranges of density contrasts for different methods for their determination is due to differences in the operation of the algorithms. This graph is similar to the spatial power spectrum, but it is easier to build. The relative frequency of this density contrast at each redshift is more obvious than in the case of the power spectrum.

\begin{figure*}[]
\begin{minipage}[h]{0.4\linewidth} 
\center{\includegraphics[scale=0.5]{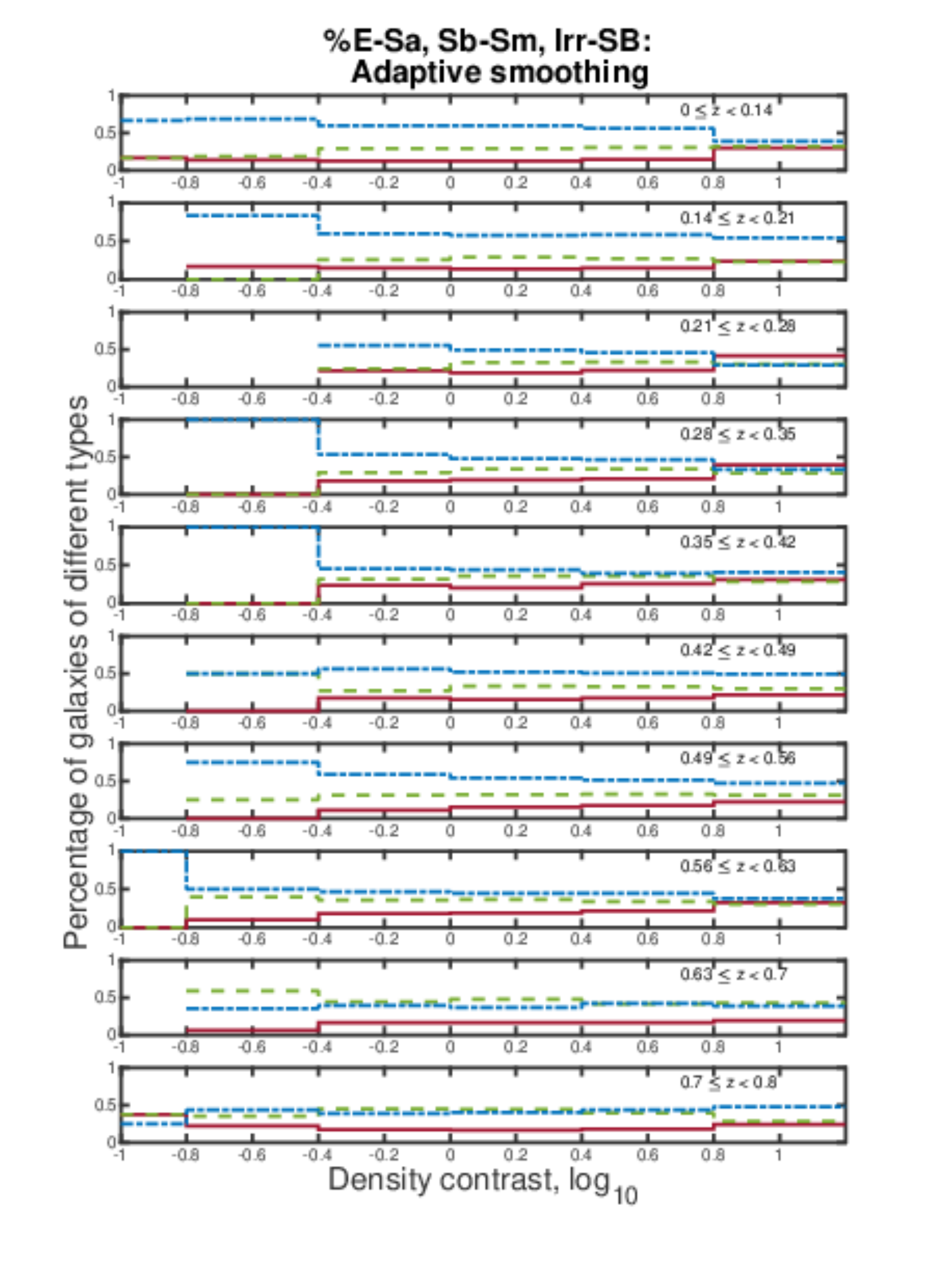}}\\  
 \end{minipage}
 \hfill
\begin{minipage}[h]{0.4\linewidth}  
\center{ \includegraphics[scale=0.5]{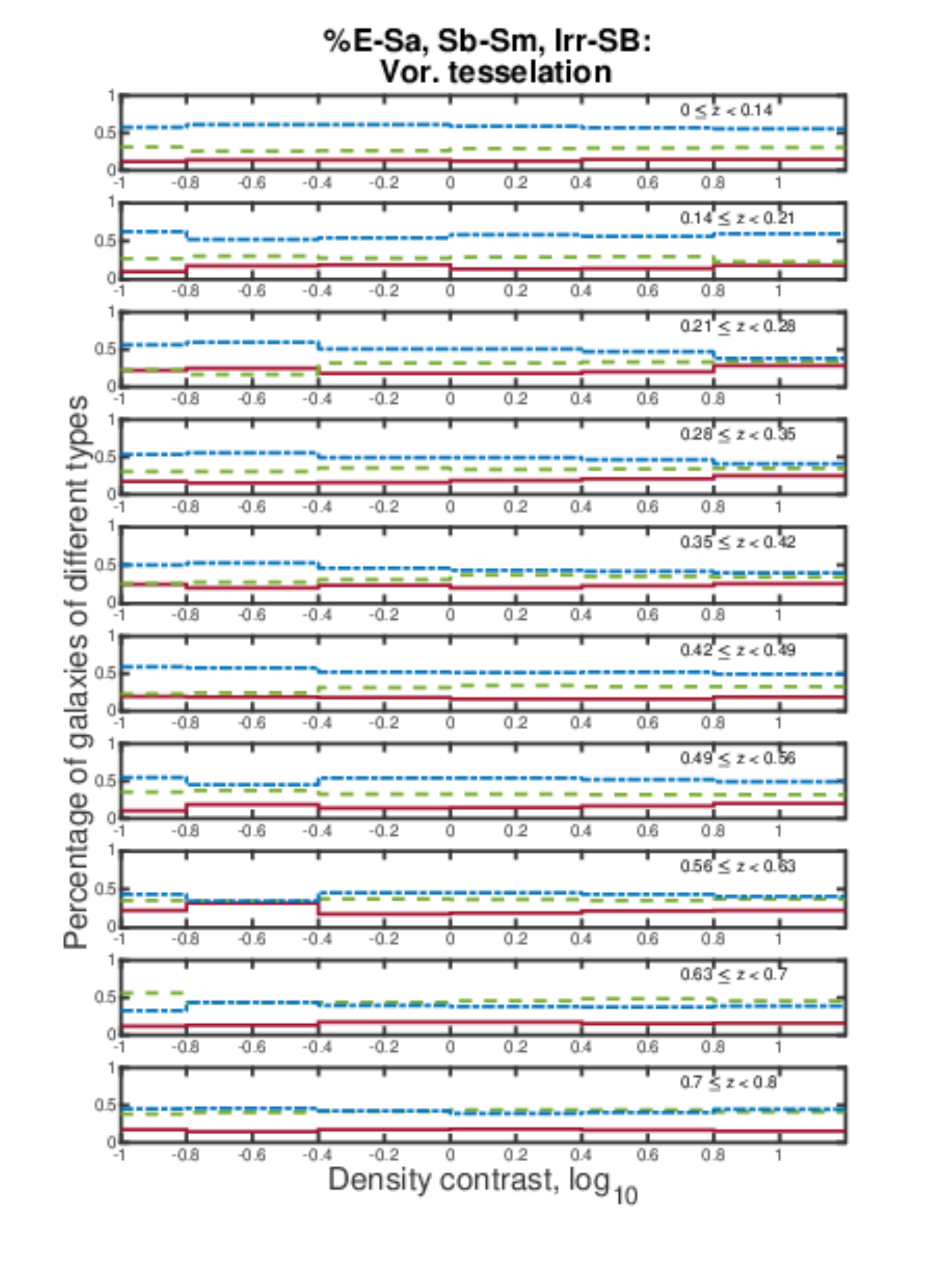}}\\  
 \end{minipage}
 \caption{The figure shows the relative number of galaxies of each type (galaxies of the early E-Sa types, galaxies of the late Sab-Sd types and irregular galaxies IRR / starburst galaxies SB; classification was carried out on the basis of the SED correspondence) in the sample under study as a function of environmental density. The full sample is divided into 10 intervals at redshift. The red solid line corresponds to the relative number of E-Sa type galaxies, the green dashed line corresponds to Sab-Sd and the blue dashed line corresponds to IRR / SB. The left panel shows the relationship obtained for the method with adaptive aperture and smoothing the density of the environment, and the right panel shows the method of Voronoi tessellation.}
\label{fig_13}
\end{figure*}

In Fig. 13 shows the correlations of the SED type of the galaxy (Section 3.2) with the density of the environment and redshift. The color and line type encoded three ranges of SED-types of galaxies:

\begin {itemize}
\item Red solid line - galaxies of early types E-Sa;
\item Green dashed line - late-type galaxies Sab-Sd;
\item Blue dashed line - irregular galaxies IRR / starburst galaxies SB.
\end {itemize}

For each redshift interval, the fraction of each range of galaxy types is shown, depending on the density of the environment. The left panel shows the relative number of galaxies for all type ranges for the density contrast obtained by the method with adaptive aperture and smoothing the density of the environment, and the right panel shows the method of Voronoi tessellation.

Numerous studies have shown a strong dependence of the relative number of early-type galaxies on the density of the environment for close redshifts (for example, \citep {Madgwick2003, Guo2013, Guo2014}). In Fig. 13 and Fig. 14 it is clearly seen that early-type galaxies prefer to dwell in denser media over the entire investigated redshift range $ 0 \leq z \leq 0.8 $. This dependence is most pronounced for the contrast of densities, determined using the method with an adaptive aperture and smoothing of the density of the environment, where the fraction of early-type galaxies sharply increases with the contrast of the density of the detected large-scale structure. This form of dependence for the method with an adaptive aperture and blurring in comparison with the method of Voronoi tessellation is due to the fact that the former identifies only significant overdensities. The total percentage of early-type E-Sa galaxies in the studied sample of galaxies systematically decreases with increasing z.

\begin{figure}[h]
\hspace*{10pt}
\includegraphics[width=1.0\linewidth]{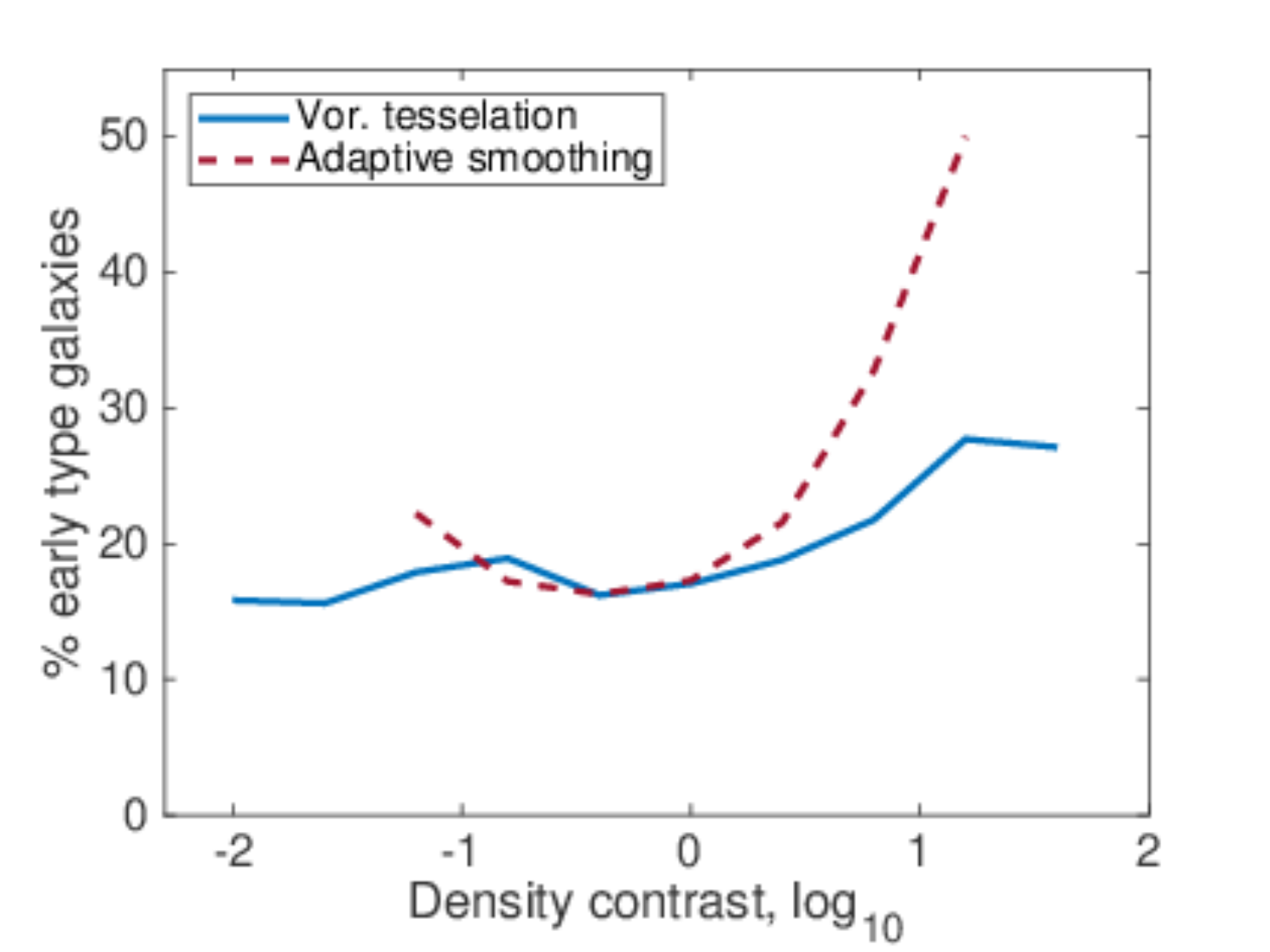}
\caption{The relative number of early-type galaxies is shown (according to the SED classification of E-Sa galaxies) depending on the contrast of the density of the environment for two methods for reconstructing density contrast maps: the method with adaptive aperture and smoothing of the density of the environment (red dashed line) and the method of Voronoi tessellation (blue solid line).}
\label{fig_14}
\end{figure}

\section {CONCLUSION}
In the course of the work, the photometric properties of a sample of galaxies of the field HS 47.5-22 were investigated (the total field area is $2.386  \; \mathrm {deg ^ 2} $). Observational data for the photometric catalog of galaxies were obtained at the 1-meter Schmidt telescope of the BAO NAS. The total sample that meets the selection criteria consists of 28,398 galaxies. For the sample, the spectral types of galaxies were determined and photometric redshifts were obtained with an accuracy of $ \sigma_z <0.01 $, which allows us to determine whether the galaxy belongs to a cluster or group.

Based on high-precision photometric redshifts for a sample of 28,398 galaxies in the redshift range up to $ \sim 0.8 $, density contrast maps were obtained, which made it possible to distinguish more than 250 significant large-scale overdensities. The resulting structures have a diverse form: from almost symmetrical circular to extended filaments. The sizes of detected density bundles are in the range from 0.5 to 10 Mpc at a comoving scale.

The quality of the detection of large-scale density clusters was estimated from 44 known clusters of galaxies in the field HS 47.5-22. 42 clusters were found, with redshifts that are in good agreement within the measurement error with data from other authors.

Density contrast distribution maps were constructed using two independent methods for determining contrast in determining density contrast in thin sections of large-scale distribution of galaxies in the redshift range up to $ \sim 0.8 $: a method with adaptive aperture and smoothing of the density of the environment and Voronoi tessellation. Both methods show consistent results, which gives us confidence in the results.

In addition, we analyzed the relationship between the types of galaxies determined by SED and the density of the environment. The results show that early-type galaxies prefer to be located in denser regions up to $ z \sim 0.8 $, which is in agreement with the earlier works \citep {Oemler1974, Dressler1980}, as well as with the results of studying the COSMOS \citep {Scoville2013} field. A complete set of density contrast distribution maps in 57 thin redshift layers, as well as layer animation, are available at \href {https://github.com/ale-gro/density_maps} {https://github.com/ale-gro/ densitymaps}.

\section*{Acknowledgements}
This work was carried out with the financial support of the
Russian Science Foundation, project 17-12-01335 ``Ionized gas in
galactic discs and beyond the optical radius''.

The authors are grateful to the referee for constructive feedback
on the initial text of the paper. 


 \end{document}